\documentclass{basi}
\usepackage{graphicx}
\def\lsim{\lower.5ex\hbox{$\; \buildrel < \over \sim \;$}}
\def\gsim{\lower.5ex\hbox{$\; \buildrel > \over \sim \;$}}
\def \simeq{\lower.3ex\hbox{$\; \buildrel \sim \over - \;$}}
\def\ch{\lower-0.55ex\hbox{--}\kern-0.55em{\lower0.15ex\hbox{$h$}}}
\def\lh{\lower-0.55ex\hbox{--}\kern-0.55em{\lower0.15ex\hbox{$\lambda$}}}

.tfm

\begin{document}

\title [Spectral and Timing evolution of GRO J1655-40 during its outburst of 2005]
{Spectral and Timing evolution of GRO J1655-40 during its outburst of 2005}

\author [D. Debnath, Sandip K. Chakrabarti, A. Nandi, and S. Mandal]
{D. Debnath$^{1,2}$, Sandip K. Chakrabarti$^{3,1,2}$, A. Nandi$^{1,4}$ and S. Mandal$^1$\\
$^1$ Indian Centre for Space Physics, Chalantika 43, Garia Station Rd., Kolkata, 700084, India\\
$^2$ Visitor, Abdus Salam International Centre for Theoretical Physics, Strada Costiera, 34100, Trieste\\  
$^3$ S. N. Bose National Center for Basic Sciences, JD-Block, Salt Lake, Kolkata, 700098, India\\
$^4$ On deputation from ISRO HQ, New BEL Road, Bangalore, 560231, India\\
e-mail: dipak@csp.res.in, chakraba@bose.res.in, anuj@csp.res.in, samir@csp.res.in}

\maketitle

\begin{abstract}
\noindent {In a recent outburst which lasted for $260$ days,
the black hole candidate GRO J1655-40 exhibited a behaviour 
similar to its last outburst observed almost eight years ago. 
We analyze a total of $150$ observational spells in $122$ days of data spreaded 
over the entire outburst phase of Feb. 2005 to Oct. 2005. From our
study, a comprehensive understanding of the detailed behaviour of this black hole candidate has 
emerged. Based on the degree of importance of the black 
body and the power-law components we divide the entire episode in four spectral
states, namely, {\it hard, soft, very soft} and {\it intermediate}. 
Quasi-Periodic oscillations (QPOs) were found in two out of these 
four states, namely, in the hard and the intermediate states. 
In the hard state, at the rising phase of the outburst, QPO frequency ranged 
from $0.034$ - $17.78$Hz and the spectra was fitted by a disk black body,
power-law and iron emission line at $6.2$ - $6.5$ keV. In the intermediate state, QPOs vary
from $13.17$Hz to $19.04$Hz and the QPO frequency modulation in this state was 
not significant. The spectra in this state are well fitted by the disk black body and the
power-law components. In the hard state of the declining phase of the outburst, 
we found QPOs of decreasing frequency from $13.14$ Hz
to $0.034$ Hz. The spectra of this state were fitted by a disk black body and 
power-law components, but in the initial few days a cooler Comptonized component
was required for a better fit. In the soft and very soft states, the spectral
states are mostly dominated by the strong disk black body component. }

\end{abstract}



\section{INTRODUCTION}

The Galactic black hole candidates are the most fascinating objects to study in X-rays,
as these sources undergo peculiar timing and spectral changes during their transient as well as
the persistent phases. The soft X-ray transient GRO J1655-40 was first observed 
by BATSE on board CGRO on 27th July 1994 (Zhang et al. 1994). This source was 
extensively observed with RXTE during 1996 and 1997 and it showed a very 
complex timing and spectral behaviour and the source was X-ray active at least for $16$ months.

GRO J1655-40, an enigmatic Low Mass X-ray Binary (LMXB) system is located at $(l,b) = 
(344.98^\circ,2.45^\circ)$ (Bailyn et. al. 1995) with R.A.=$16^h54^m00^s$ and 
Dec.=$-39^\circ50^m 45^s$. Its mass (M = $7.02\pm0.22~M_\odot$; Orosz \& Bailyn 1997) 
distance (D = $3.2\pm0.2$~kpc; Hjellming \& Rupen 1995), 
and inclination angle ($\theta = 69.5^\circ\pm0.1^\circ$; Orosz \& Bailyn 1997) are well determined. 
The mass of its companion star is = $2.3~M_\odot$ (Bailyn et. al. 1995). GRO J1655-40
may also have shown signatures of the ejection of the superluminal radio jet (Tingay et. al. 1995; 
Hjellming \& Rupen 1995). The maximum speed of the jet was found to be $\sim 0.37 c$. 
Recent VLT-UVES spectroscopic observations suggest that the distance 
to the source is $\le 1.7$ kpc (Foellmi et. al., 2006) with a secondary star of spectral type of F6IV,
making it one of the closest known black hole candidates.

The first observed outburst showed a double peaked profile in the ASM light curve and it is 
quite different from other black hole candidates. The first peak in May, 1996, the source
showed a strong flaring activity with non-thermal emission, whereas during the second peak in
August, 1997, the source spectrum was softer and thermal, except near the end of the outburst
when its spectrum was hard (Sobczak et. al. 1999). At least three distinct spectral states, 
namely, very high state, high/soft state and low/hard state (Sobczak et al. 1999)
have been reported. The luminosity variation of the outburst was of fast rise and exponential 
decay (Chen et. al. 1997). Investigation of X-ray timing properties of GRO J1655-40 during 
the 1996-97 outburst revealed QPOs varying from $0.1$ Hz to $300$ Hz (Remillard et. al. 1999). 
Two very important discoveries  were found there: one is the superluminal radio jet 
(Tingay et. al. 1995; Hjellming \& Rupen 1995) and the other is the existence of very high 
QPO frequencies ($300$ \& $450$ Hz) (Remillard et. al. 1999, Strohmayer 2001).

After remaining `dormant' for almost eight years, GRO J1655-40 showed a renewed X-ray activity
in the late February 2005 (Markwardt \& Swank 2005, Chakrabarti et al. 2005,
Shaposhnikov et al. 2007). The source remained active in X-rays for the next $260$ days
and during this period and it was extensively observed with the RXTE Satellite. In the present  paper,
we analyze the archival data of RXTE instruments (ASM and PCA) and present the results for both 
the timing \& the spectral properties of GRO J1655-40 during this outburst phase. 
In the entire outburst phase, we identified four spectral states 
characterized by the presence or absence of a soft black body component at low energy and 
the power-law component at higher energies above $\sim 10$ keV. Since there are
confusions in the literature regarding the nomenclature vis-{\'a}-vis the properties,
we define them here at appropriate places. The four identified states 
are termed as the hard, soft/very-soft, intermediate and hard states. During the total outburst 
we observed the transitions in this sequence: hard $\rightarrow$ soft/very soft $\rightarrow$ 
intermediate $\rightarrow$ hard. In each of these spectral states, we carried out the 
timing analysis and find QPO frequencies. The Power Density 
Spectra (PDS) are quite different in different spectral states and sometimes the 
nature of the PDS is highly correlated with the spectral features. The justification
of these four classifications will be presented later. In previous communications 
(Chakrabarti et al. 2005; Chakrabarti et al. 2008) the evolution of 
the QPO frequencies with time was shown in the initial and final outburst stage. The 
rapid variation in QPO frequencies was explained by using an oscillating and propagating shock.

Prior to our present analysis, Shaposnikov et. al. (2007) carried out a multi-wavelength 
study for the early stage (beginning with 21st of February, 2005) of the outburst of
GRO J1655-40 for a total of 25 days of data using instruments like
RXTE \& INTEGRAL for X-rays, VLA for radio study and ROSTE \& SMARTS for optical region. 
On the basis of their multi-wavelength campaign they classified the spectral states of the observed 
period in four spectral states, namely, low-hard, hard intermediate, soft intermediate, high-soft.
After correlating X-ray and radio fluxes they concluded that the physical origins of the radio
emission and the X-ray emission are not the same. The evidence of a closer coupling 
between the power-law component and QPO as also observed by Vignarca et. al. (2003)
is totally consistent with the shock propagation model of Chakrabarti et al (2005, 2008)
as the shock does not propagate in the {\it disk} as they mentioned, but through the sub-Keplerian flow
which surrounds the disk (e.g., Chakrabarti \& Titarchuk, 1995). 

Our study, on the other hand, covers 122 days of the observational data spreading 
over the full period of the outburst. On the basis of the results of RXTE data, we classified 
the total outburst in a slightly different way with four distinct spectral states.
Furthermore, we thoroughly studied the QPO behaviour. We got QPOs in a
total of $67$ observations out of a total of $150$ observations. We also studied the photon count
variation in different energy bands for different spectral states via hardness and
softness intensity diagrams. We identify the energy band
in which QPOs are predominantly seen. We show spectral components
and their flux variations. We claim that two components of the flow, namely, the 
Keplerian and the sub-Keplerian (halo) are necessary to explain the mass accretion dynamics. 
We theoretically estimate the disk and the halo rates from spectral fits of several observations. 

In passing, we may mention that some other workers reported analysis of the outburst using Swift 
(Brocksopp et al. 2006) and XMM-Newton \& INTEGRAL (Trigo et al. 2007).
The Suzaku data of the late phase of the outburst has been analyzed by Takahashi et al. (2008)  who
showed that two different Comptonizing electron clouds are required to explain the 
high energy spectra in the low/hard state. This agrees with our findings also (Chakrabarti et al. 2008). This
will be illustrated in more detail in the present paper as well. 

The paper is organized in the following way: In the next Section, we analyze
the data and present the results of our analysis. This includes the timing analysis 
of ASM and PCA data and spectral analysis of PCA data. In Section 3, we 
present the brief interpretation of the overall results. Finally, in Section 
4, we make concluding remarks.

\section {Observation and Data Analysis}

\subsection {Analysis of ASM and Light Curves}

We analyze publicly available observational data from the RXTE instruments of the 2005 outburst. 
Here, we present the results from the All Sky Monitor (ASM) and Proportional Counter Array (PCA) 
covering the entire eight months of the outburst of GRO J1655-40. Our analysis covers from 
the 25$^{th}$ of February, 2005 (MJD = 53426) to 16$^{th}$ of October, 2005 (MJD = 53659). 
The ASM data has four energy bands corresponding to $2-3$ keV, $3-5$ keV, $5-12$ keV 
and $2-12$ keV. PCA contains five proportional counter units (PCUs 0-4). We used only 
PCU 2 data for both the timing and spectral analysis due to its reliability and it 
is on for $100\%$ of the goodtime. Data reduction and analysis were carried out with 
the FTOOLS version of HEADAS-6.1.1 software and XSPEC version 12.3.0.

\begin{figure}

\vbox{
\vskip -2.0cm
\centerline{
\includegraphics[scale=0.6,angle=0,width=14truecm]{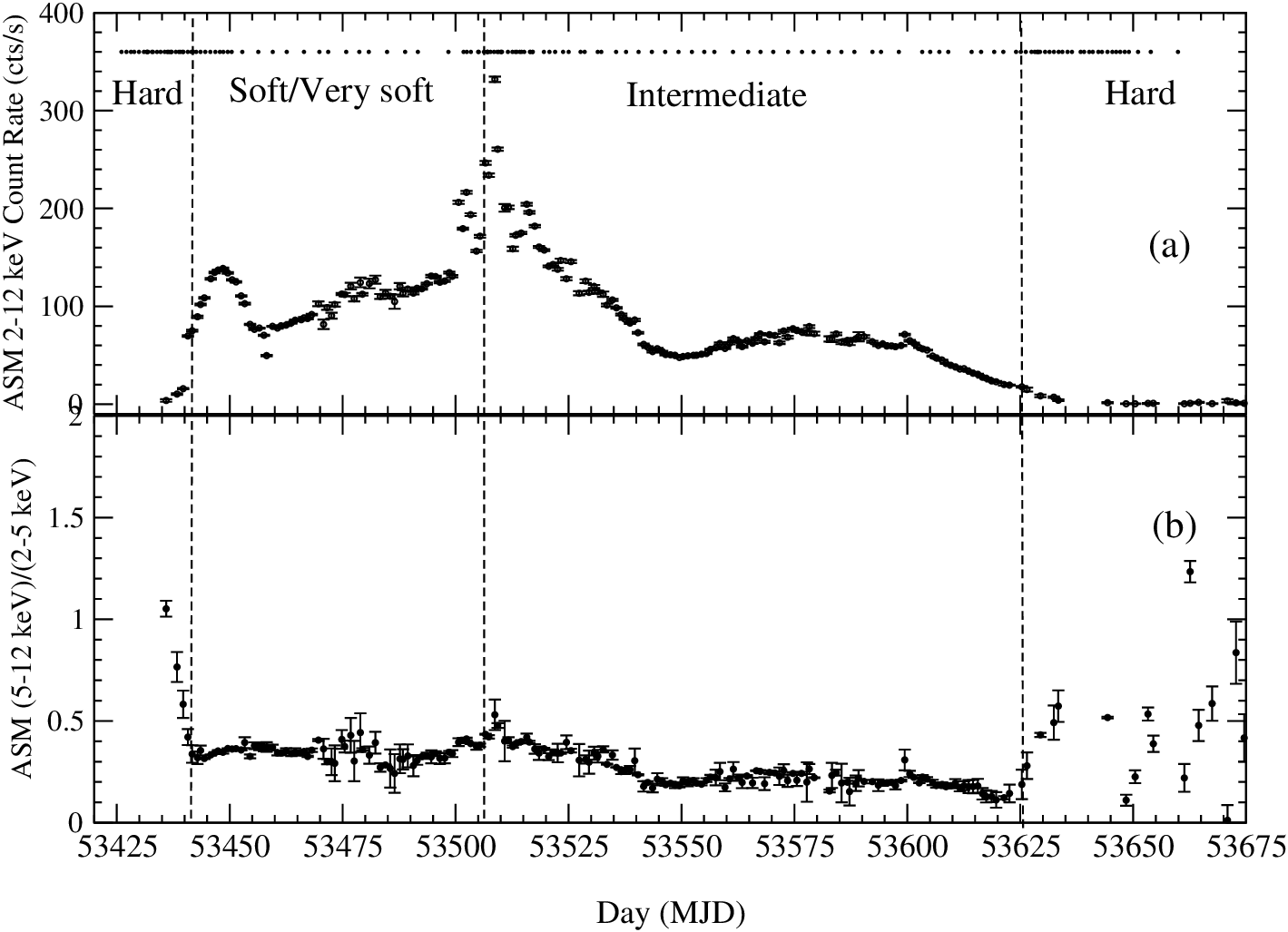}}
\vspace{0.0cm}
\noindent{\small {\bf Figure 1:} (a) 2-12 keV ASM light curve and (b) hardness ratio 
(5-12 keV vs. 2-5 keV count ratio) as a function of the MJD of the event.
The vertical dashed lines indicate the transition of states.}}
\end{figure}

We have extracted and analyzed the ASM (Levine et al. 1998) data of different energy bands for 
the entire observation. In Figs. 1(a-b), the total 2-12 keV ASM light curve (counts/sec) and the  
ASM hardness ratio (ratio of the photon count rates in 5-12 keV  and 2-5 keV bands) are
plotted. The origin of the time axis is MJD 53420 ($19^{th}$ February, 2005),
which is six days before the initial rise of the X-ray intensity. The hardness ratio
variation distinctly reflects the state transitions. The hard to soft transition 
takes place on the $13^{th}$ of March, 2005 (MJD = 53442), the soft to intermediate 
transition on the $16^{th}$ of May, 2005 (MJD = 53506), and the intermediate to hard 
transition takes place on the $12^{th}$ of September, 2005 (MJD = 53625). These are
marked on the plot. However, the local changes in the spectral features of different states 
are not evident from this plot. This leads us to conduct a robust spectral 
analysis using the PCA data and the results are presented below.

\begin{figure}
\vbox{
\vskip 0.0cm
\centerline{
\includegraphics[scale=0.6,angle=0,width=8truecm]{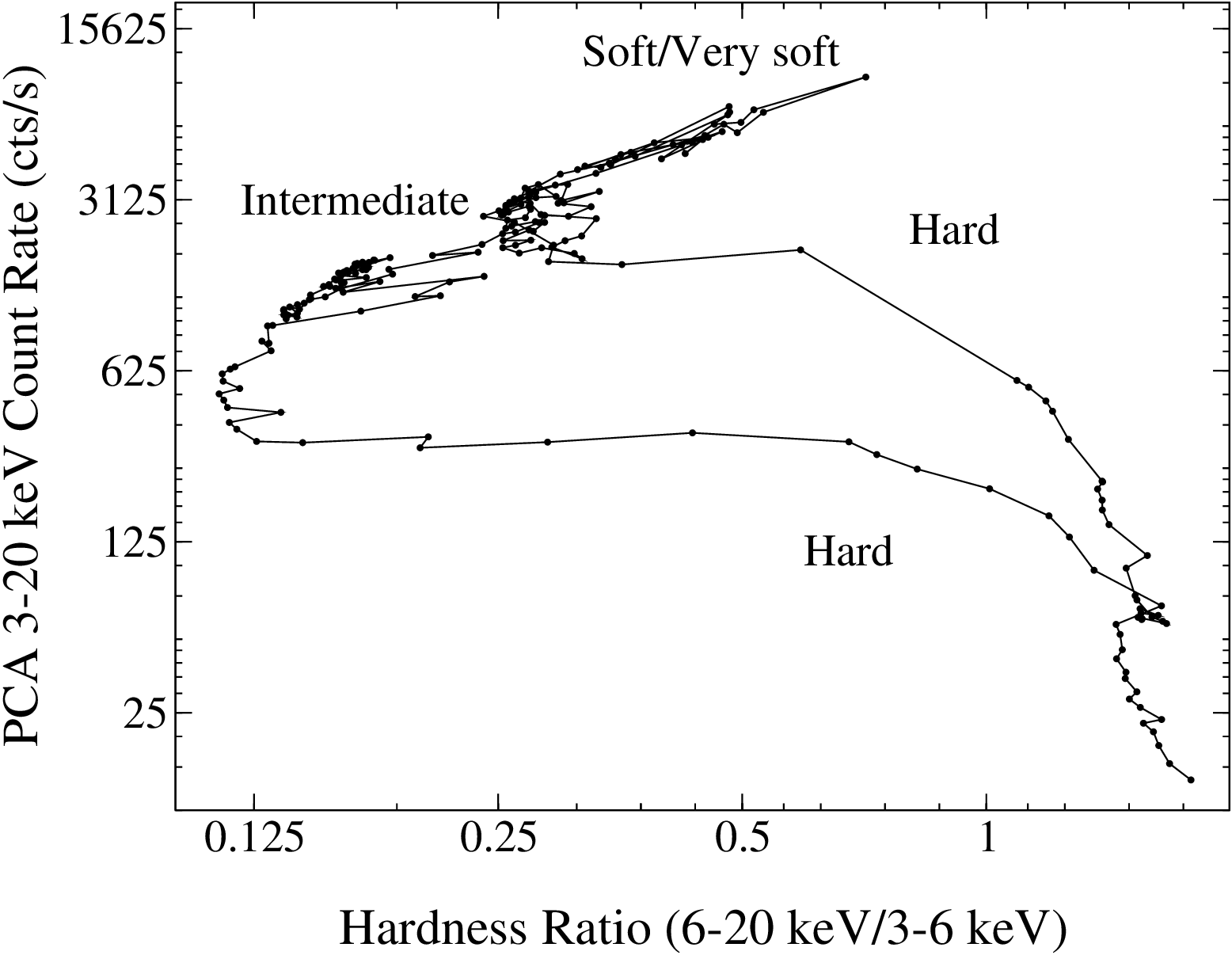}}
\vspace{0.0cm}
\noindent {\small {\bf Figure 2:} The Hardness Intensity Diagram (HID) 
observed with RXTE/PCA. Count rates are in 3-20 keV energy band and 
hardness ratio is defined as the ratio of count rates in the 6-20 keV 
and 3-6 keV bands.}}
\end{figure}

The RXTE archival data from February 25$^{th}$, 2005 (MJD = 53426) to October 16$^{th}$, 2005 (MJD = 53659) 
were extracted and analyzed from the Proportional Counter Array (PCA; Jahoda et al., 1996). We extract
light curves (LC), PDS (with $0.01$s binning of PCA data from $3-25$keV) 
and the energy spectra from the good and the best-calibrated detector units 
{\it i.e.}, PCU2, for the PCA. We use the latest FTOOLS software package.
For the timing analysis (LC \& PDS) from February $25^{th}$, 2005 (MJD = 53426) to March, $11^{th}$, 
2005 (MJD = 53440), we use the Science Data of the Event mode ($E\_125us\_64M\_0\_1s$, FS4f*gz) and 
for the rest of the observed dates we use the Science Data of the Binned mode 
($B\_8ms\_16A\_0\_35\_H$, FS37*.gz) and of the Event mode ($E\_62us\_32M\_36\_1s$, FS3b*.gz). 
To extract the light curves from the Event 
mode data files, we use the ``sefilter" task and for the Binned mode data files, 
we use the ``saextrct" task. For the spectral analysis, we use Standard2 
Science Data of PCA (FS4a*.gz). The ``pcabackest" task was used for the PCA background estimation purpose. 
Here we used bright source epoch5 background model file for calculating PCA background. 
We also incorporated $pca\_saa\_history$ file for taking care of saa data. 
To generate the response files, we use the ``pcarsp" task. For the rebinning of the pha 
files created by the ``saextrct" task, we use the ``rbnpha" task. 

In Fig. 2 we plotted the PCA 3-20 keV count rate of the 2005 outburst against X-ray color 
(ASM count ratio between 6-20 keV and 3-6 keV energy bands).
It is evident that the pre and post-outburst phases tend to appear and disappear
from the low count region having harder spectrum. In GRO J1655-40 outburst,
the transition occurs from the spectral states {\it hard $\rightarrow$ soft (very soft) 
$\rightarrow$ intermediate $\rightarrow$ hard}. It is observed that the rapid changes 
in the hardness ratio occurs only in the hard states, whereas in the soft and intermediate states
the hardness ratio changes very slowly. Both rising \& falling arms of the diagram 
corresponds to the hard state. In both the cases, we found the presence of QPOs. The 
possible physical origin will be discussed in Section 2.

In the first phase of the hard state from the 25$^{th}$ of February, 2005 (MJD = 53426) to 
the 12$^{th}$ of March, 2005 (MJD = 53441), we found QPOs from $34$ mHz to $17.78$ Hz. 
The observed QPO frequencies were found to be increased monotonically with time 
(day) from 0.082 Hz to 17.78 Hz (on the first day another QPO at $34$mHz was also seen.)
The soft state starts from the 13$^{th}$ of March, 2005 (MJD = 53442) and
continued till 15$^{th}$ of May, 2005 (MJD = 53505). In this region no QPO was observed. 
The intermediate state is seen from the 16$^{th}$ of May, 2005 (MJD = 53506) to 11$^{th}$ 
of September, 2005 (MJD = 53624). Interestingly, we found QPOs only for $8$ days, 
from 16$^{th}$ of May, 2005 (MJD = 53506) to 20$^{th}$ of May, 2005 (MJD = 53510) 
and from 25$^{th}$ of May, 2005 (MJD = 53515) to 27$^{th}$ of May, 2005 (MJD = 53517).
In between, for four days we observed no signature of QPOs. The QPO frequencies 
varied from $13.17$ Hz to $19.04$ Hz. In the PDS, we also found one broad QPO bump 
at frequencies near $7$ Hz. The final hard state observed is from the 
12$^{th}$ of September, 2005 (MJD = 53625) to 16$^{th}$ of October, 2005 (MJD = 53659). 
The QPOs of $0.023$ Hz to $20.20$ Hz QPOs were observed in this state. If we follow
one of the QPO frequencies, we find it to decrease monotonically from $13.14$ Hz 
to $0.034$ Hz within $20$ days.

\begin{figure}
\vbox{
\vskip 0.0cm
\centerline{
    \includegraphics[scale=0.5,angle=0,width=8truecm]{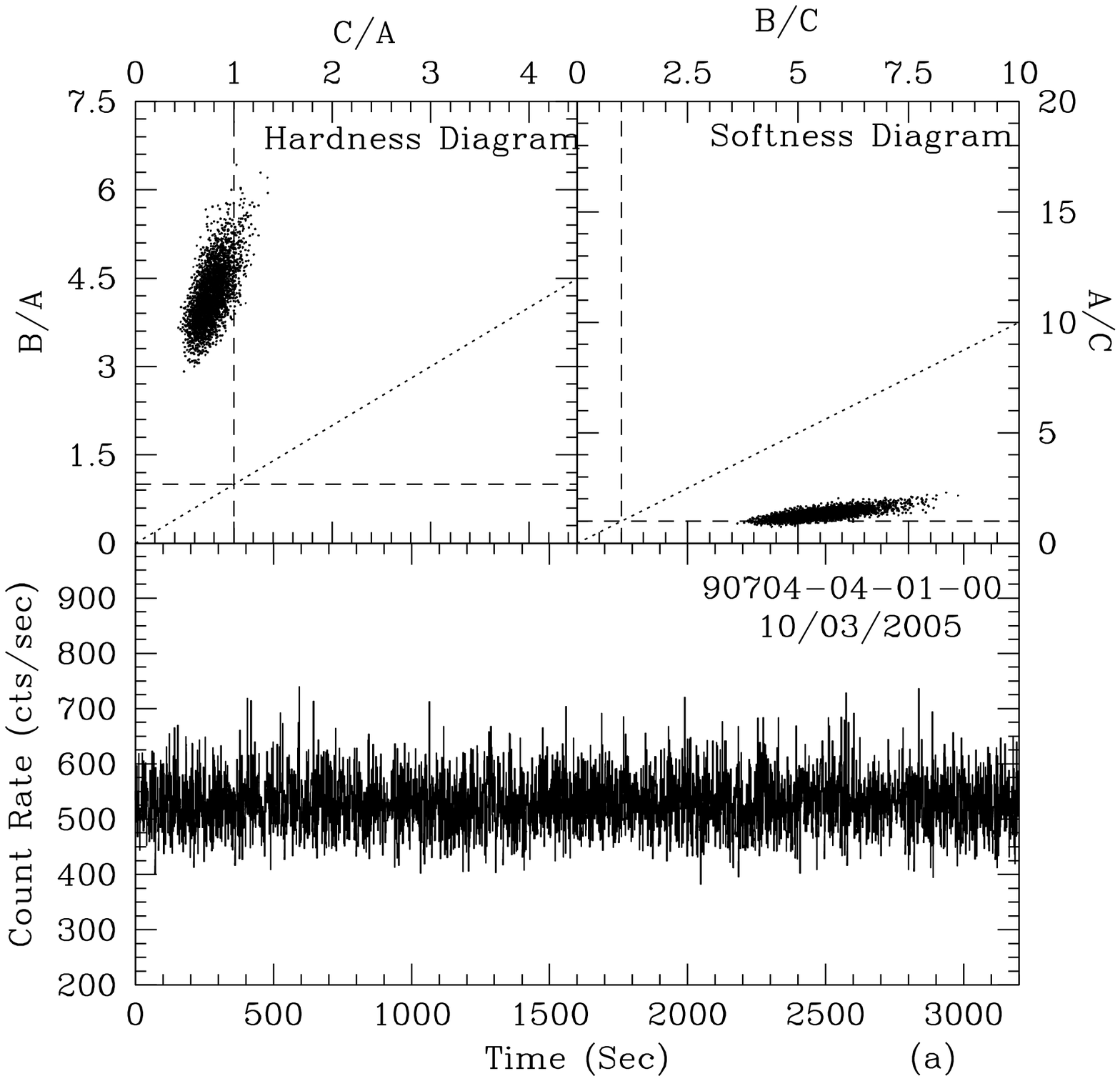}
    \includegraphics[scale=0.5,angle=0,width=8truecm]{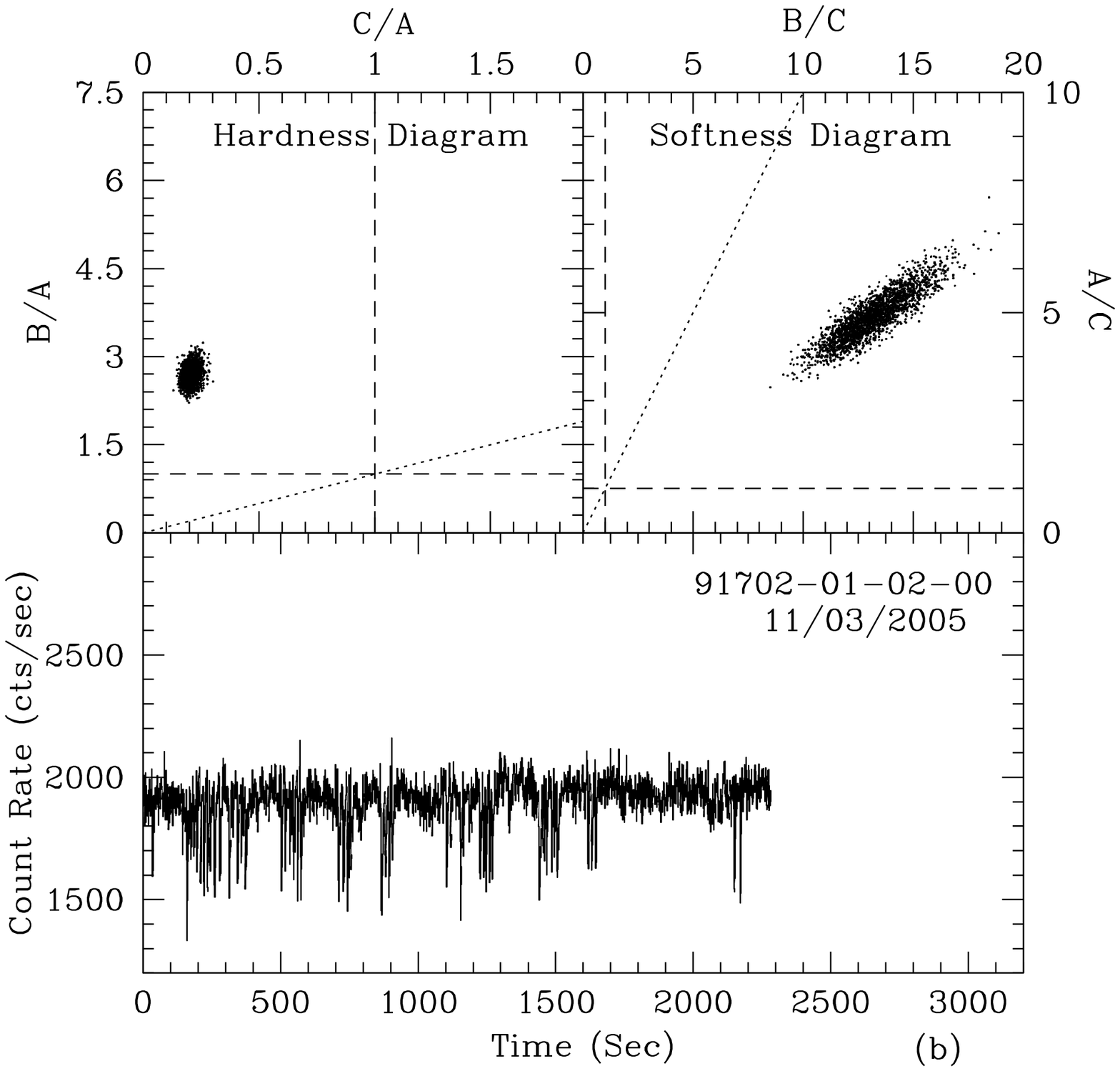}}
\vspace{0.0cm}
\noindent {\small {\bf Figure 3 (a-b):} In the lower panel, 2 - 15 keV (0-35 Channels) PCA light 
curve and in the upper panel the hardness and softness diagrams are plotted. Both the figures are of 
hard state observed on 10th of March and 11th of March, 2005. Drastic changes in timing 
features are observed in these two consecutive days, see text for details. In hardness diagrams,
the dashed horizontal ($B=A$), vertical ($C=A$) and the dotted line ($B=C$) are for 
reference purpose. In softness diagram they represent $C=A$, $B=C$ and $A=B$
respectively.}}
\end{figure}

\begin{figure}
\vbox{
\vskip 0.0cm
\centerline{
   \includegraphics[scale=0.5,angle=-0,width=8truecm]{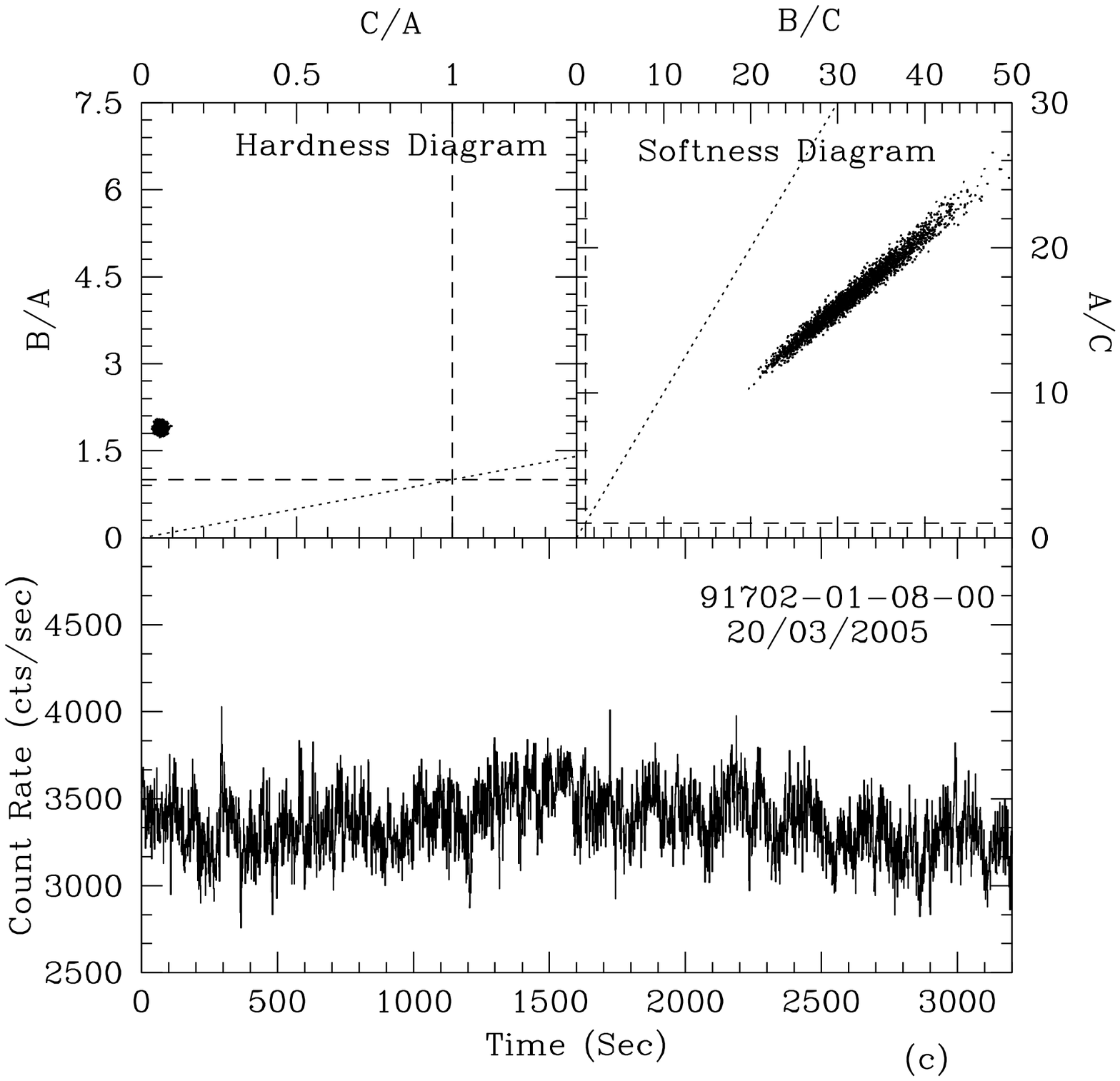}}
\vspace{0.0cm}
\noindent {\small {\bf Figure 3 (c):} Same as in Figs. 3(a-b), except that
the observation is of 20th March, 2005 when the source was in a soft state.}}
\end{figure}

\subsection {\bf Timing Analysis}

We carried out the detailed timing analysis of the total $122$ days data of $150$ observational IDs. 
We used the PCU2 data from the Event mode ($E\_125us\_64M\_0\_1s$) and Science Array mode 
($B\_8ms\_16A\_0\_35\_H$) data for the timing analysis. Our timing analysis is mainly to 
study the light curves with hardness and softness variation and the PDS of each data. 
Out of these observations, we find QPOs in a total of $67$ observations made in $43$ days.
A summary of the results are presented in Table 1. Here, we list the observing date and 
time and the PCA count rates  (photon counts/sec) for PCU2 in 3 different 
energy bands, E1: $2-3.5$ keV ($0-7$ channels), 
E2: $3.5-10.5$ keV ($8-24$ channels) and E3: $10.5-60$ keV ($25-138$ channels). We also list 
two hardness ratios (E2/E1 and E2/E3) and the observed QPO frequencies (in Hz).\\

\subsubsection {\bf Light curves with hardness and softness diagrams}

We extracted 2 - 15 keV (0-35 Channels) PCA light curve with a time bin of 1 sec. To have the 
qualitative analysis of photon count variations in different energy bands, we plotted both the 
hardness and softness ratio variations. To plot the hardness and the softness ratios, we extracted light 
curves for three energy bands: $A:0-8$ channels ($2-4$ keV), $B:9-35$ channels ($4-15$ keV) and 
$C:36-138$ channels ($15-60$ keV). A hardness diagram is the plot between $C/A$ vs. $B/A$
while the softness diagram is the plot between $B/C$ vs. $A/C$. Our motivation of splitting 
the energies in this way stems from the fact that the Keplerian disk primarily emits
at a low energy ($\lsim 4$ Kev) for the mass of the black hole we are interested in. 
Thus, $A$ will be emitted mostly from the Keplerian component. The component $B$ would be 
emitted from the region where the moderate thermal Comptonization of the Keplerian photons 
take place. The component $C$ would be emitted from the region which 
is definitely depleted or enhanced during state transitions as it represents
the higher energy side of the pivotal energy [$\sim 15$ keV] in the spectrum. 
Thus, these diagrams are not directly connected to the spectral states -- 
rather, they are connected to the geometry, i.e., the number of soft photons produced
by the Keplerian disk ($\sim A$) and the seed photons intercepted by the 
`Compton cloud' [$\sim (B + C)$] and the number of scatterings they undergo 
($\sim B$ or $\sim C$). 

\noindent (a) Hard State in the rising phase:

In Figs. 3(a-e), we plot the hardness and softness diagrams along with the light 
curves in the days when the source exhibited different spectral states. In 
Fig. 3a and Fig. 3b, the light curves belong to the hard state in the rising phase 
of the outburst and are of the 10$^{th}$ and the 11$^{th}$ of March, 2005. On 
the 11$^{th}$ of March, 2005, the photon count is several times than that of 
the previous day. Though all the three components increased, there is a 
drastic change in the hardness and the softness ratio diagrams because $C$ 
was increasing much slower than $A$ and $B$. Thus the spectrum become
much softer within one day. Strong QPO features were observed in both the 
days. The detailed PDS and spectral features are discussed later below.
Interestingly, as will be shown below, on the 11th March of 2005, the source did 
not exhibit any QPO feature in low energy X-ray (2-4 keV), but it is present 
in the observation on the 10th of March. Finally, on the 12th March of 2005, 
the source enters into soft state and the QPO is totally absent. 
On both the days $B> A > C$.

\noindent (b) Soft State in the rising phase:

In Fig. 3c, we draw a similar Figure with the data of the 20th of 
March, 2005, when the source was in the soft state. The $C$ component
is further reduced while $A$ and $B$ continue to go up with $A$ approaching $B$.
No QPO signature is observed in this state. In this case $B>A >>C$. 

\begin{figure}
\vbox{
\vskip 2.0cm
\centerline{
    \includegraphics[scale=0.5,angle=0,width=8truecm]{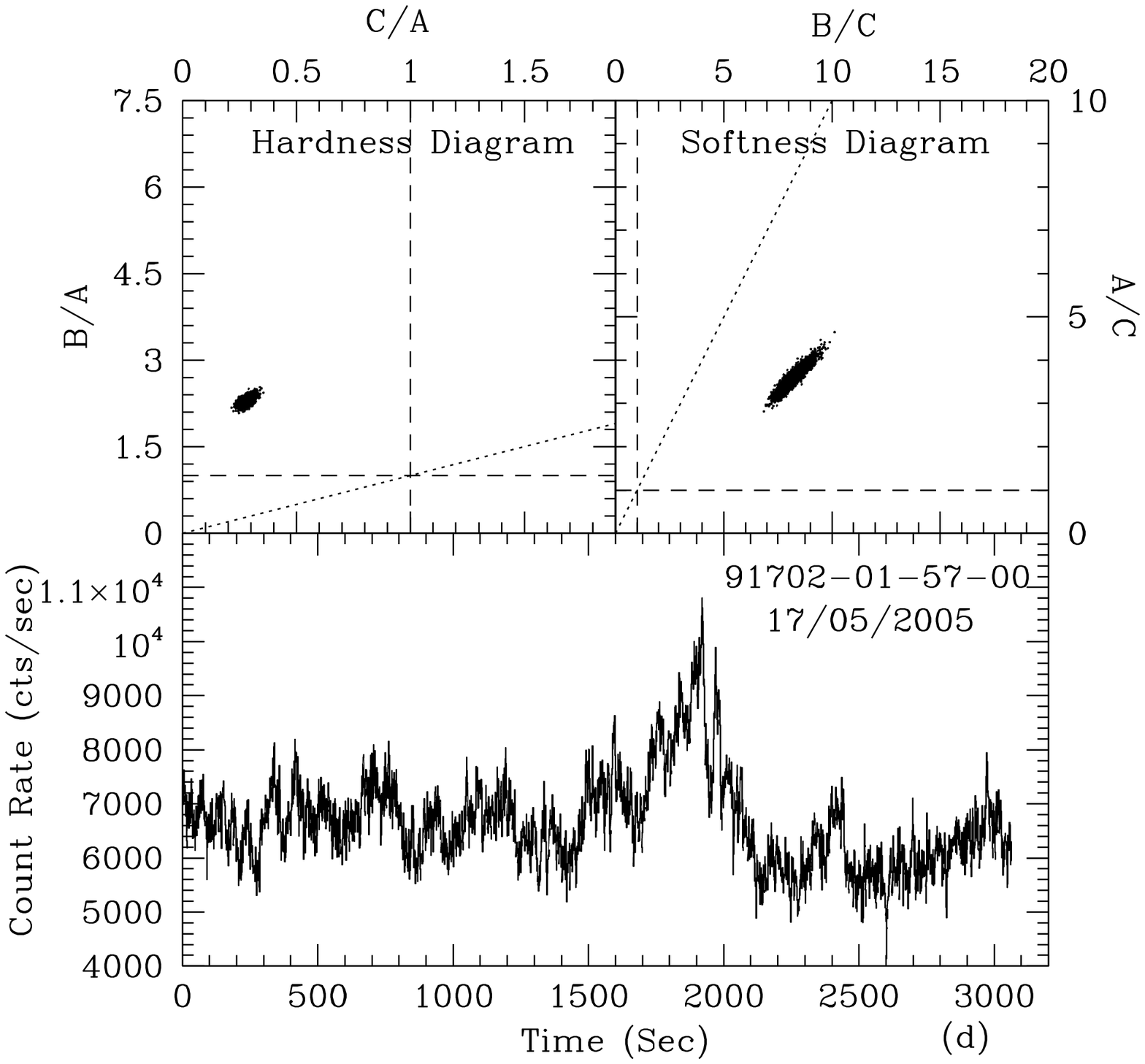}
    \includegraphics[scale=0.5,angle=0,width=8truecm]{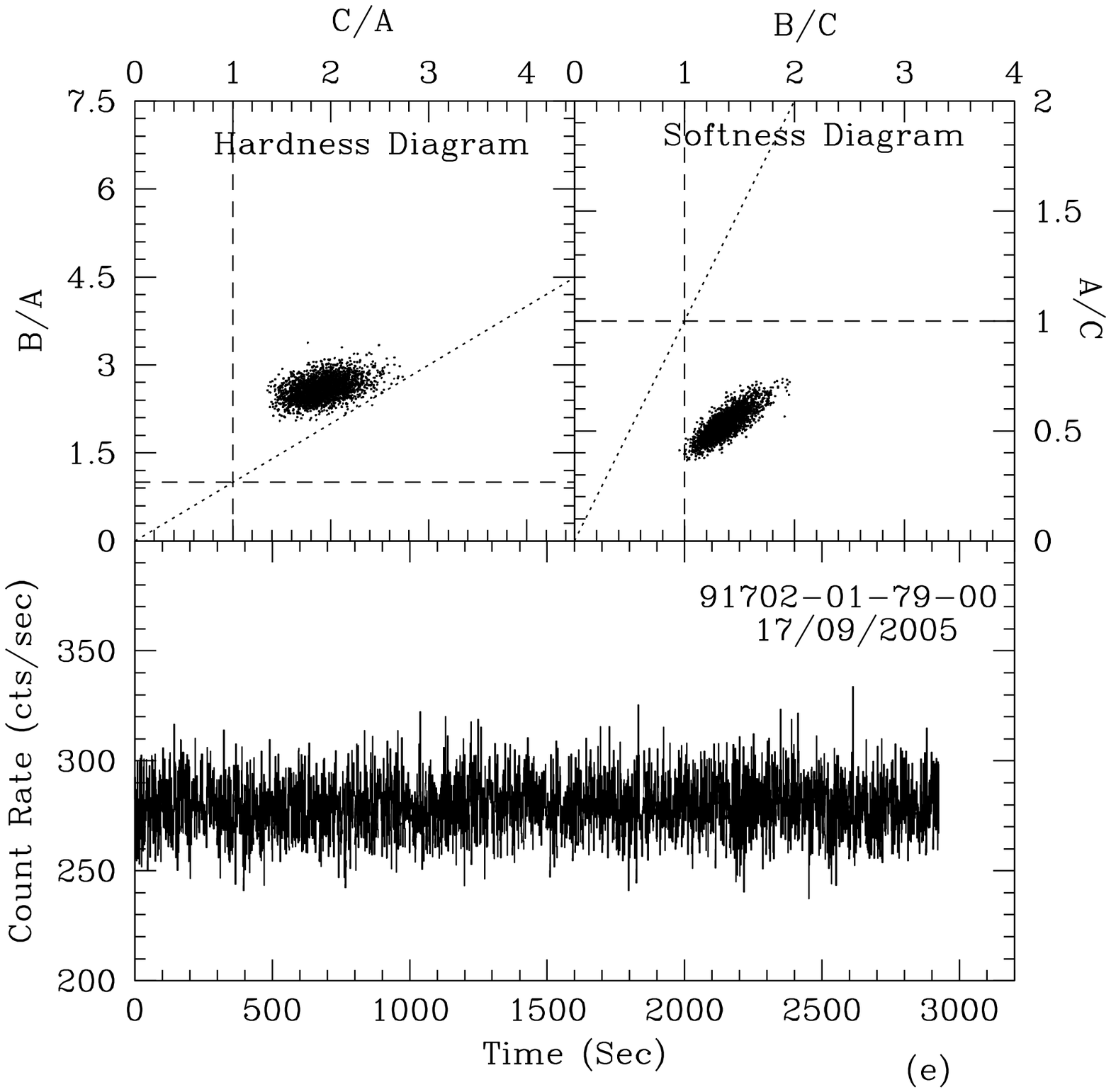}}
\vspace{0.0cm}
\noindent {\small {\bf Figure 3 (d-e):} Same as in Figs. 3(a-b) except for (d) the intermediate state 
(17th of May, 2005) and (e) the hard state of the decline phase (17th of Sept., 2005).}}
\end{figure}

\noindent (c) Intermediate State:

In Fig. 3d, we present the light curve and the hardness/softness diagram in the intermediate 
state as observed in a typical day (17th May, 2005). The $C$ component is increased very rapidly 
while the others increasing very slowly. This state shows some evidence of QPOs on certain days.
Here also $B>A>C$.

\noindent (d) Hard State in the decline phase:

Finally, in Fig. 3e, when the source is in the hard state again, both the slopes of the hardness 
and the softness diagrams are more flat as compared to those in the intermediate state
and have a similar characteristics as that of the hard state. A typical case 
based on the observation on the 17$^{th}$ of September, 2005 is shown here. In this case, 
the components $B$ and $C$ became dominant are more than the photon counts in $A$. Here, $B > C > A$. 
The tendencies of the ratios $C/A$ and $B/A$ in relation to the count rate are consistent with that
shown in Figs. 3(a-b). Again, QPO has started appearing in this state and the frequency went down 
as the days progressed.

\subsubsection {\bf Power Density Spectra}

Figures 4-8 show results of the PDS for the entire episode. 
To generate Power Density Spectrum (PDS), we have used "powspec" task of XRONOS package with a
normalization factor of `-2' to have the 'white' noise subtracted rms fractional
variability. The power obtained has the unit of rms$^2$/Hz.
The light curve of X-ray variability from which PDS was obtained were binned 
at $0.01$ sec time resolution so that the Nyquist frequency is $50$ Hz. 
QPOs are generally Lorentzian type (Nowak 2000, van der Klis 2005) and thus
each PDS was fitted with
a power-law plus Lorentzian profile to derive the central frequencies and widths of 
each observed QPO. One has to be careful in rebinning the frequency scale (Papadakis and Lawrence 1993) as 
it may misrepresent the behaviour especially at low frequencies. In our case, 
we rebinned the PDS with a geometrical factor of $-1.02$ to have a nearly equispaced log(frequency) bin.
For  this choice, any QPO below $0.0122$Hz would not be detected.
For the best fitting of the PDS as well as QPO profiles we
used the least square fit technique. After fitting PDS, we have used "fit err" 
task to calculate +/- error for QPO frequencies and widths. This task calculates
the 90\% confidence range of any fitted parameter. For the best fit
we occasionally use another broad Lorentzian component at the break frequency position. In Table 2, 
we present a summary of the results where we put the centroid frequency ($\nu$) of the 
QPO, its width ($\Delta\nu$) (both in Hz),  the coherence parameter $Q$ (= $\nu$/$\Delta\nu$). 
The RMS amplitudes ${\cal R}$ of the fitted QPOs are also included which were 
calculated from ${\cal R}= 100 (PW\pi/<\phi>)^{1/2}$, where, $P$, $W$ and $\phi$ 
are the power, half-width ($\Delta\nu/2$) of the Lorentzian fitted QPO and 
the mean count rate of the source respectively. If $Q > 2$, it is considered to be a strong QPO, 
otherwise it is not strong and look more like a bump on the PDS. Since we are 
interested only in the QPO properties, namely, the frequencies associated with the 
QPO, bump and the break, only these are included in the Table and not the power-law features
which may been used for the best fit. Since fitting the total PDS is not our goal,
an F-test is not needed to check whether the extra model components are required or not.

In Fig. 4a, we present the model fitted PDS of the light curve of $10^{th}$ March,
2005 (ID: 90704-04-01-00). We used the ``Constant + Lorentzian + Lorentzian
+ Lorentzian + Power-Law" models for the fitting. QPOs are at $2.313$Hz \& $4.599$Hz 
with a $0.363$Hz break frequency. The higher frequency QPO is clearly 
the first harmonic frequency. Index of the Power-Law after the break is $-0.383$.
In Fig. 4b, the PDS of $20^{th}$ March, 2005 (ID: 91702-01-08-00) is shown. This
is fitted with ``Power-Law + Power-Law" models. This is akin to a soft state PDS.
Index of the first Power-Law is = $-0.6339$ and the second power-Law is $-0.5767$. 
In Fig. 4c, we show the model fitted PDS in the intermediate-state  on $17^{th}$ May, 2005 
(ID: 91702-01-57-00G). We used ``Power-Law + Lorentzian + Lorentzian" models for the fitting. 
Here the QPO frequency of $18.94$ Hz with a QPO bump at frequency $7.65$Hz. 
Index of the first Power-Law is = $-1.299$. In Fig. 4d, the result of $17^{th}$ September, 
2005 (ID: 91702-01-79-00), when the object was in the hard state of the declining phase
is shown. We used ``Power-Law + Lorentzian + Lorentzian + Lorentzian + Lorentzian + Power-Law" 
models for the fitting. Here we found QPOs at frequencies of $0.203$Hz, $8.71$Hz \& $17.39$Hz 
(the last one being a higher harmonic) with a break frequency at $1.77$ Hz.  

\begin{figure}
\vbox{
\vskip -1.5cm
\centerline{
       \includegraphics[scale=0.6,angle=270,width=8truecm]{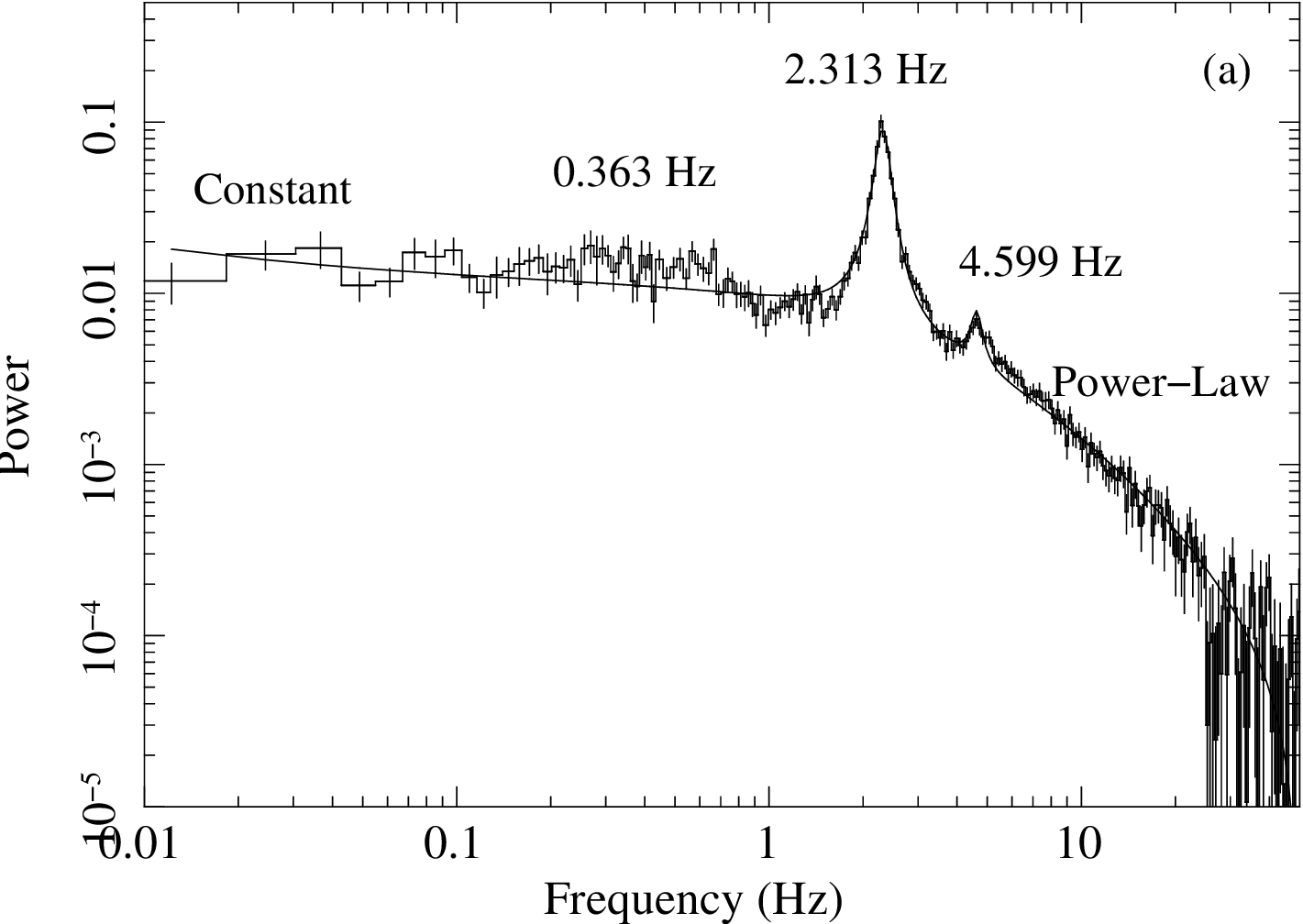}
       \includegraphics[scale=0.6,angle=270,width=8truecm]{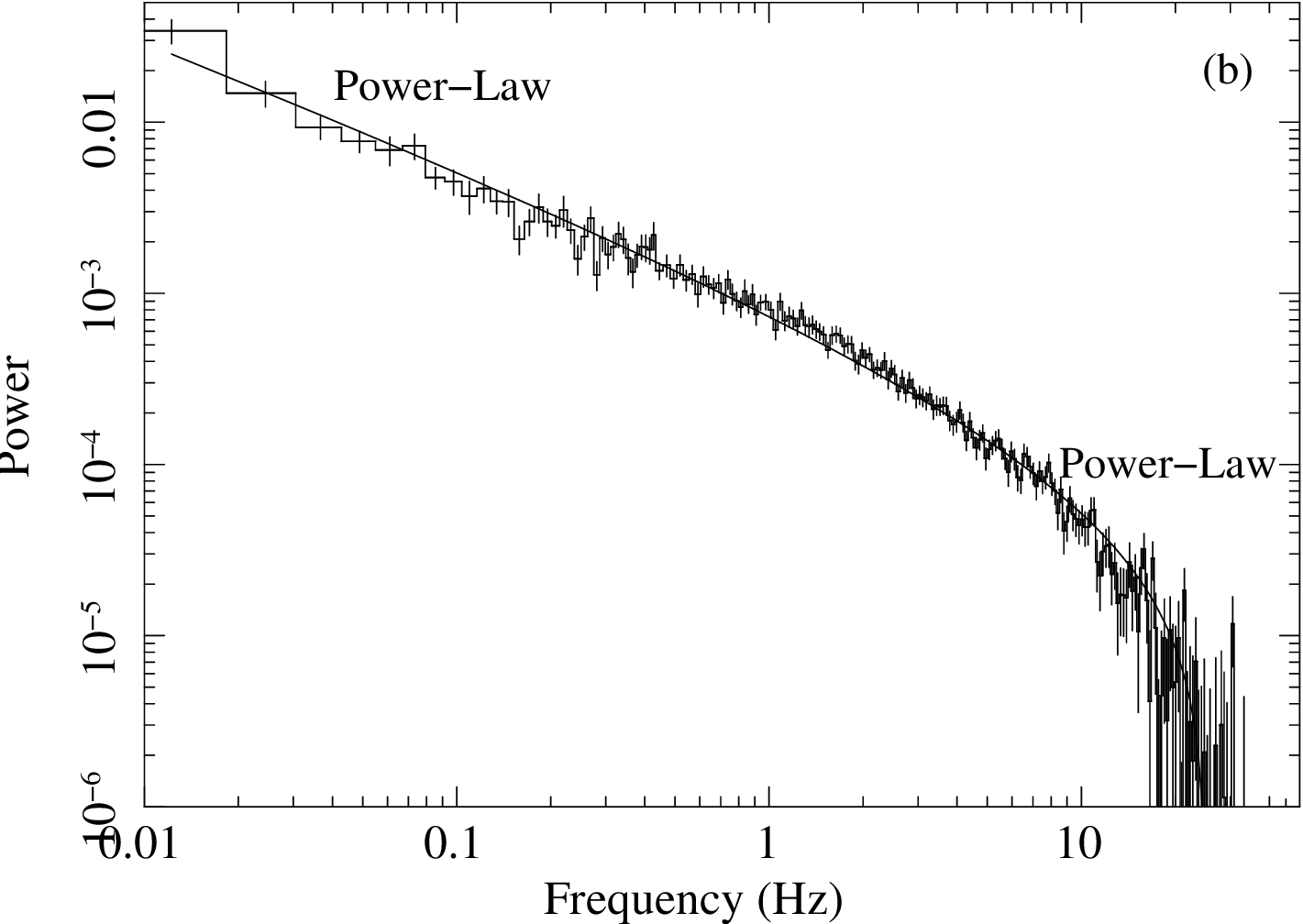}}
\vspace{0.0cm}
\noindent {\small {\bf Fig. 4(a-b):} (a) The model fitted PDS of $10^{th}$ March, 
2005 (ID: 90704-04-01-00). A QPO is found at $2.313$ Hz,  with $0.363$ Hz break frequency.
(b) The model fitted PDS of $20^{th}$ March, 2005 (ID: 91702-01-08-00). No QPO is observed in this case.}}
\end{figure}

\begin{figure}
\vbox{
\vskip -0.0cm
\centerline{
       \includegraphics[scale=0.6,angle=270,width=8truecm]{fig4c.ps}
       \includegraphics[scale=0.6,angle=270,width=8truecm]{fig4d.ps}}
\vspace{0.0cm}
\noindent {\small {\bf Fig. 4(c-d):} (c) The model fitted PDS of $17^{th}$ May, 2005 
(ID: 91702-01-57-00G). The QPO is at $18.94$Hz with a bump at $7.65$Hz.
(d) The model fitted PDS of $17^{th}$ September, 2005 (ID: 91702-01-79-00). 
QPOs are found at $0.203$Hz, $8.71$ Hz  with a break frequency at $1.77$ Hz.}}
\end{figure}

\begin{figure}
\vbox{
\vskip 0.0cm
\centerline{
       \includegraphics[scale=0.6,angle=0,width=8truecm]{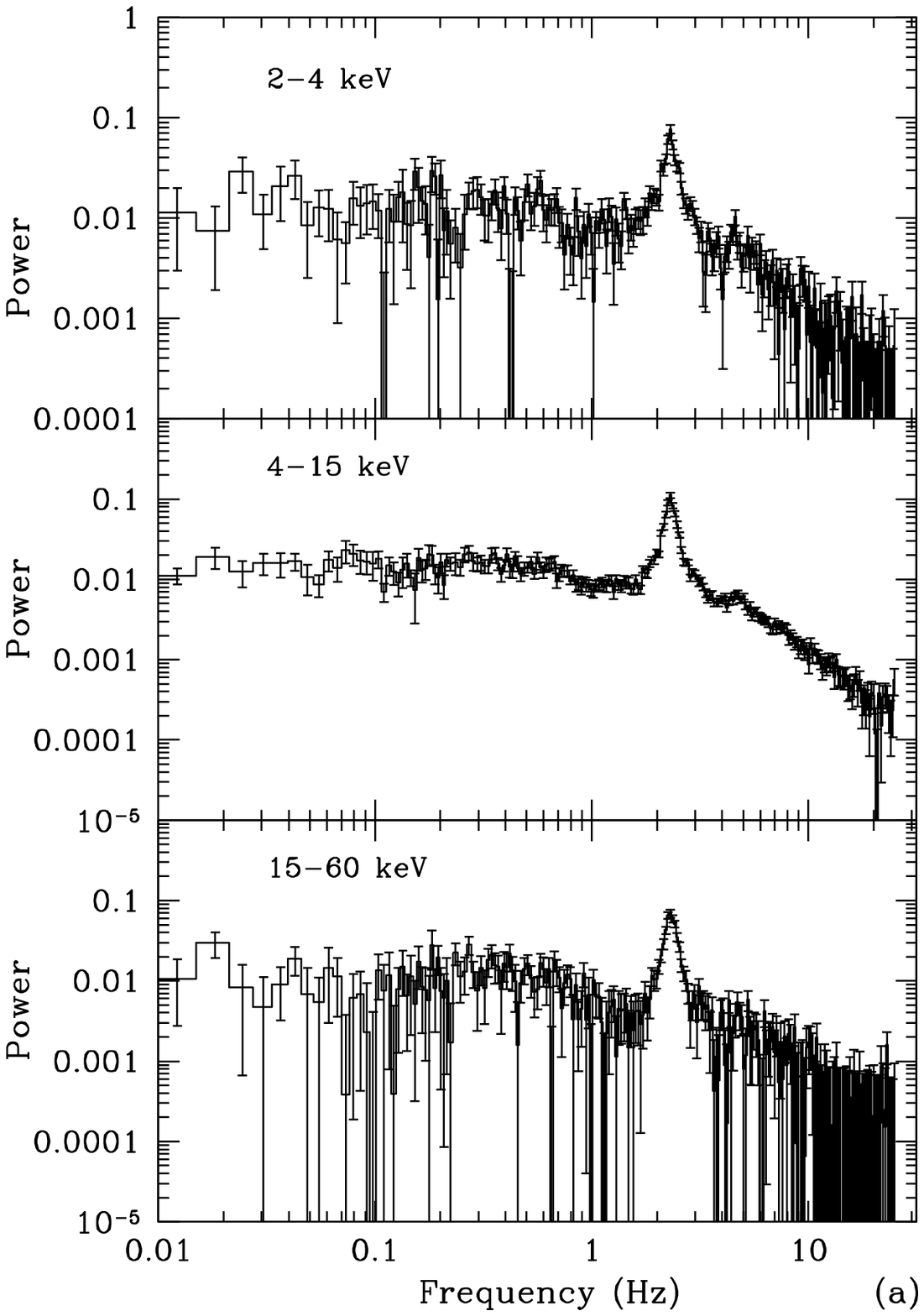}
       \includegraphics[scale=0.6,angle=0,width=8truecm]{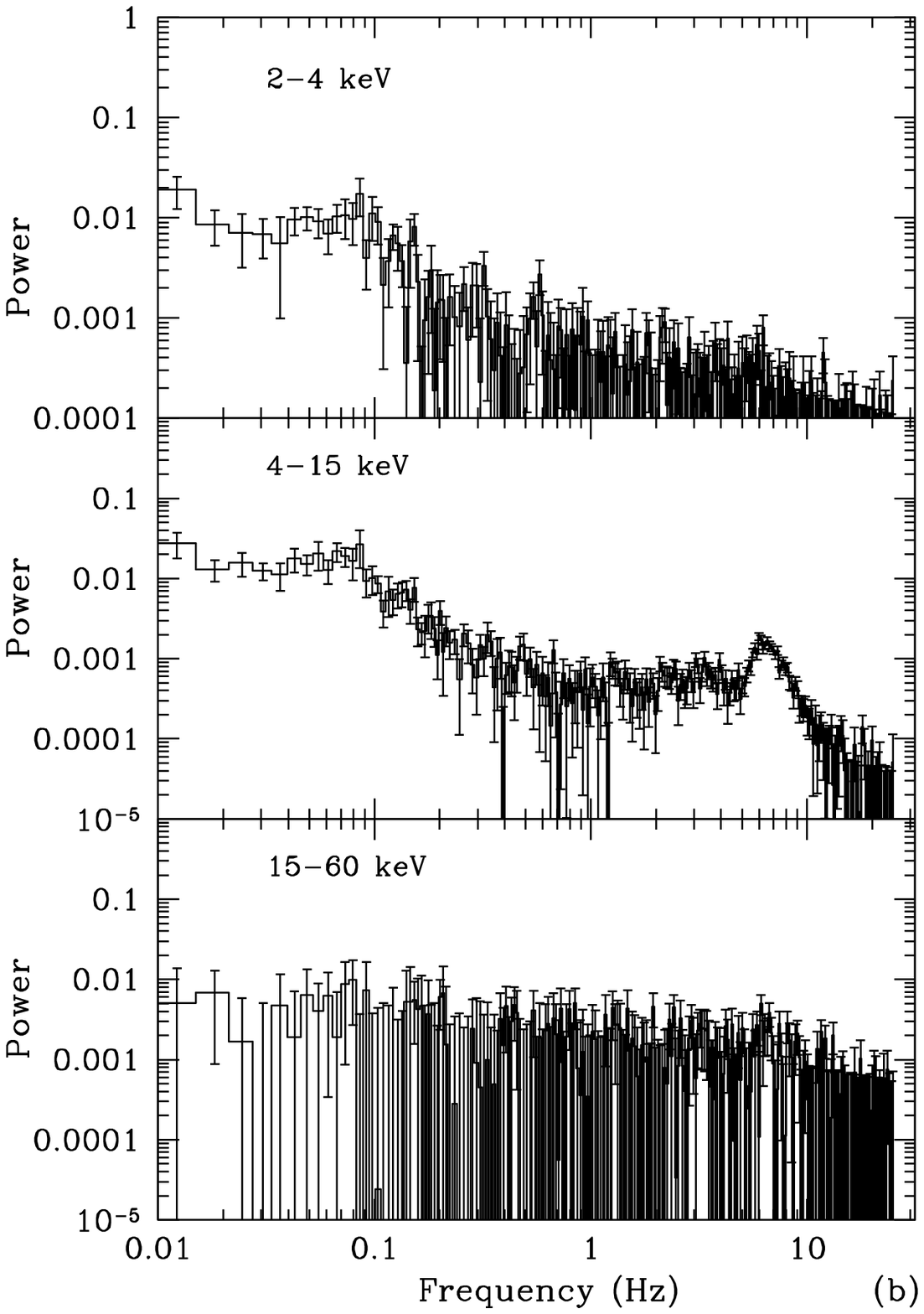}}
\vspace{0.0cm}
\noindent {\small {\bf Fig. 5(a-b):} Energy dependence of the PDS. The upper, middle and the
lower panels are for $2 - 4$ keV, $4 - 30$ keV and $30 - 60$ keV respectively. (a) Data
of  $10^{th}$ March, 2005 (ID: 90704-04-01-00). Both the soft and the medium energy X-rays
show the $2.313$ Hz QPO, the power is higher in medium energy by fifty percent. (b) Data 
of $11^{th}$ March, 2005 (ID: 91702-01-02-00G) shows that the QPO at $6.522$Hz is exhibited 
only by hard photons ($4-25$ keV).}}
\end{figure}

\begin{figure}
\vbox{
\vskip 0.0cm
\centerline{
       \includegraphics[scale=0.6,angle=0,width=8truecm]{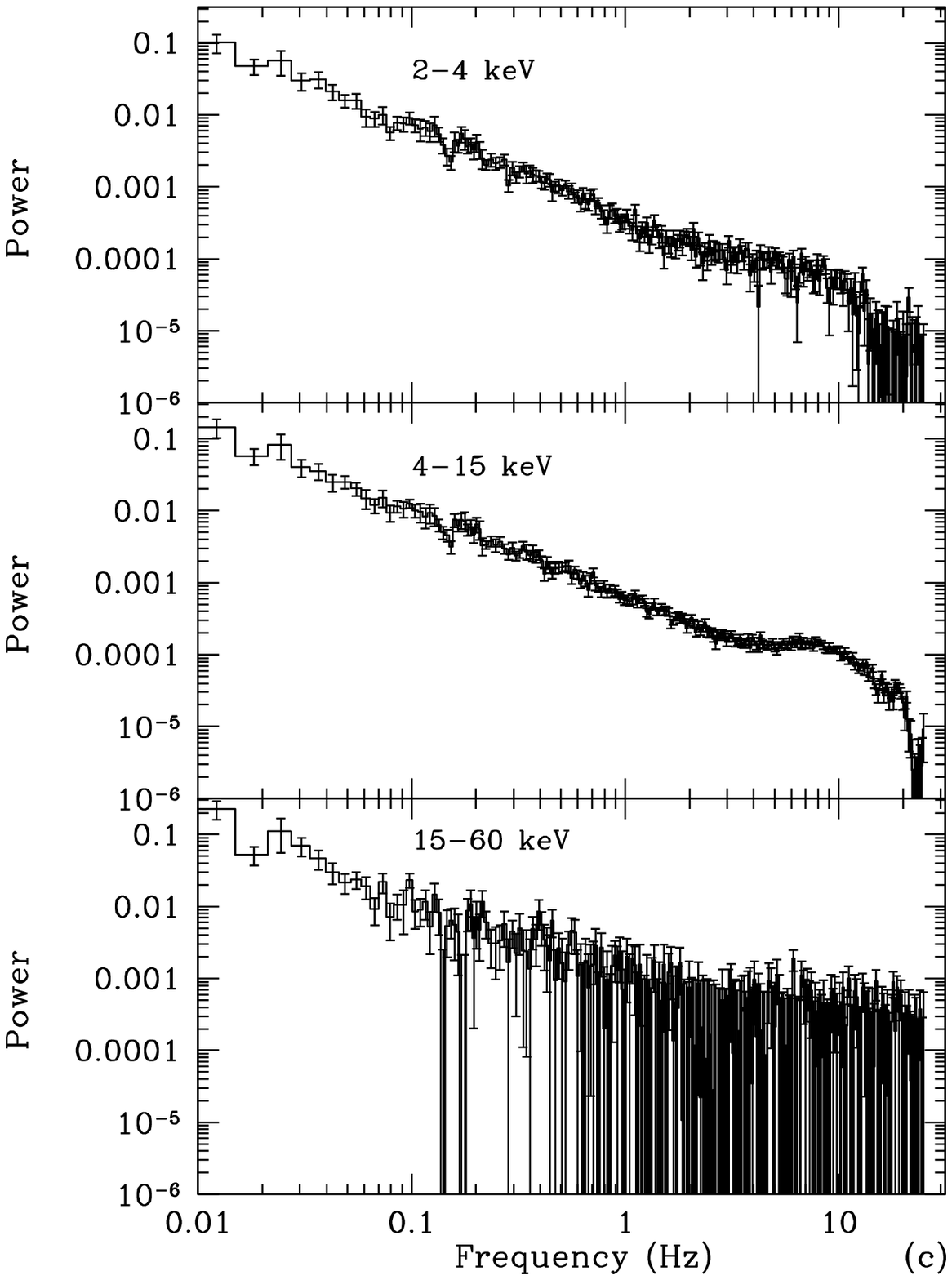}
       \includegraphics[scale=0.6,angle=0,width=8truecm]{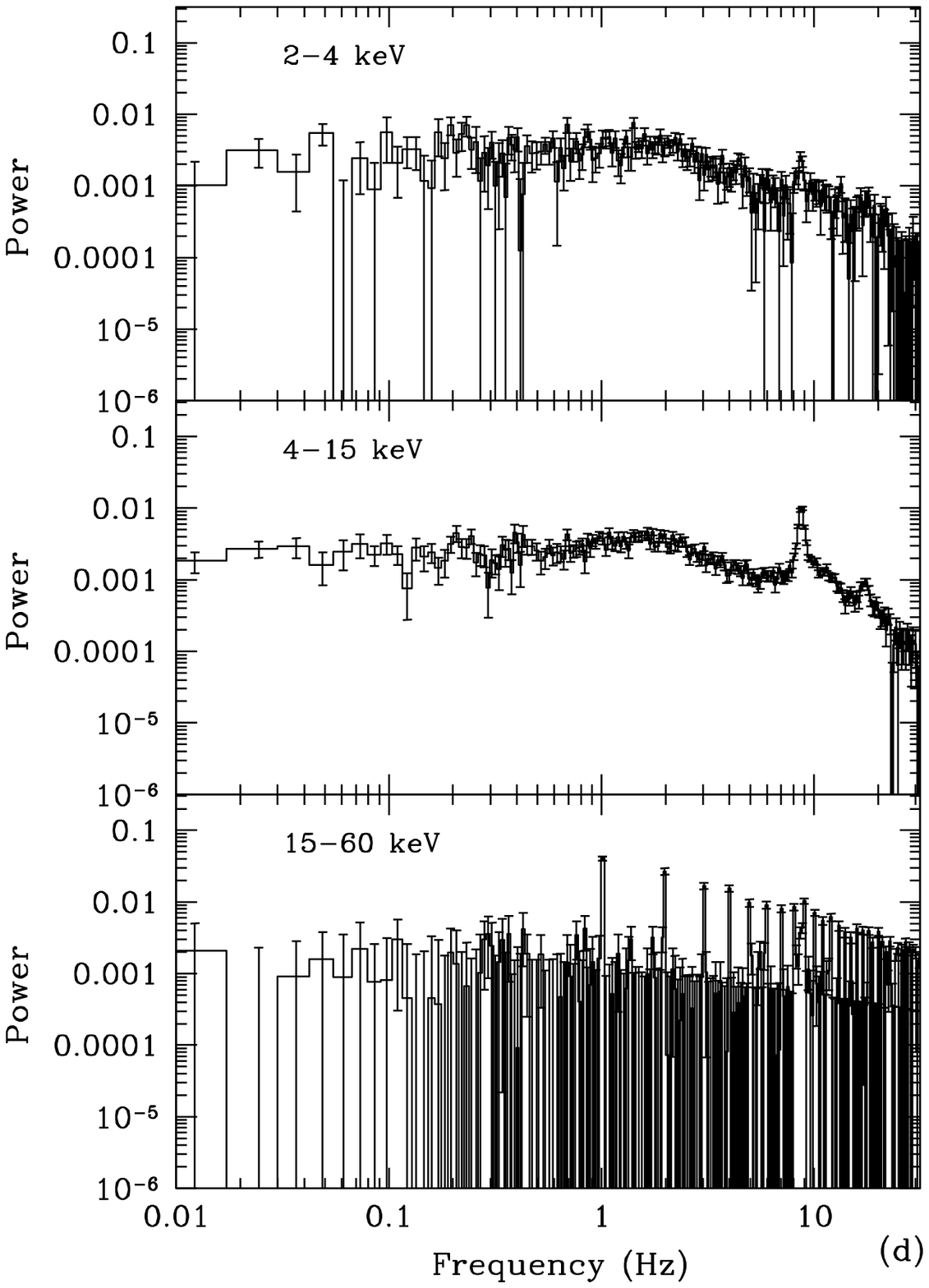}}
\vspace{0.0cm}
\noindent {\small {\bf Fig. 5(c-d):} (c) PDS of the data of $17^{th}$ May, 2005 (ID: 91702-01-57-00G) 
showing that the QPO is exhibited by hard photons (4 - 25 keV) only. (d) PDS of the data of $17^{th}$ 
September, 2005 (ID: 91702-01-79-00). Here QPO is seen in both the soft and the intermediate energies 
as in the hard state of the rising phase (a).}}
\end{figure}

The rising and the declining phases of the outburst showed a very exciting feature. The
QPO frequency increased monotonically in the rising phase, while it is decreased monotonically
in the declining phase. In Fig. 6 we present the PDS for each day in the rising phase. 
Arrows indicate the direction in which the date (marked in parenthesis as dd/mm) 
increases. The observation IDs and the QPO frequencies are also shown in the inset. 
In Chakrabarti et al. 2005, the trend of the rising phase has been discussed. 
In Fig. 7, we present the PDS variation in the intermediate state. In the inset 
we mark frequency at which the bump is formed in case QPO frequency was unavailable. 
The variation of these frequencies seem to be a bit arbitrary. On the other hand, the variation 
of PDS in the declining phase (Fig. 8) shows monotonically decreasing QPO frequency. 
We discuss the implication of these interesting observations in the next Section.
 
\begin{figure} 
\vbox{
\vskip -1.0cm
\centerline{
       \includegraphics[scale=0.6,angle=0,width=14truecm]{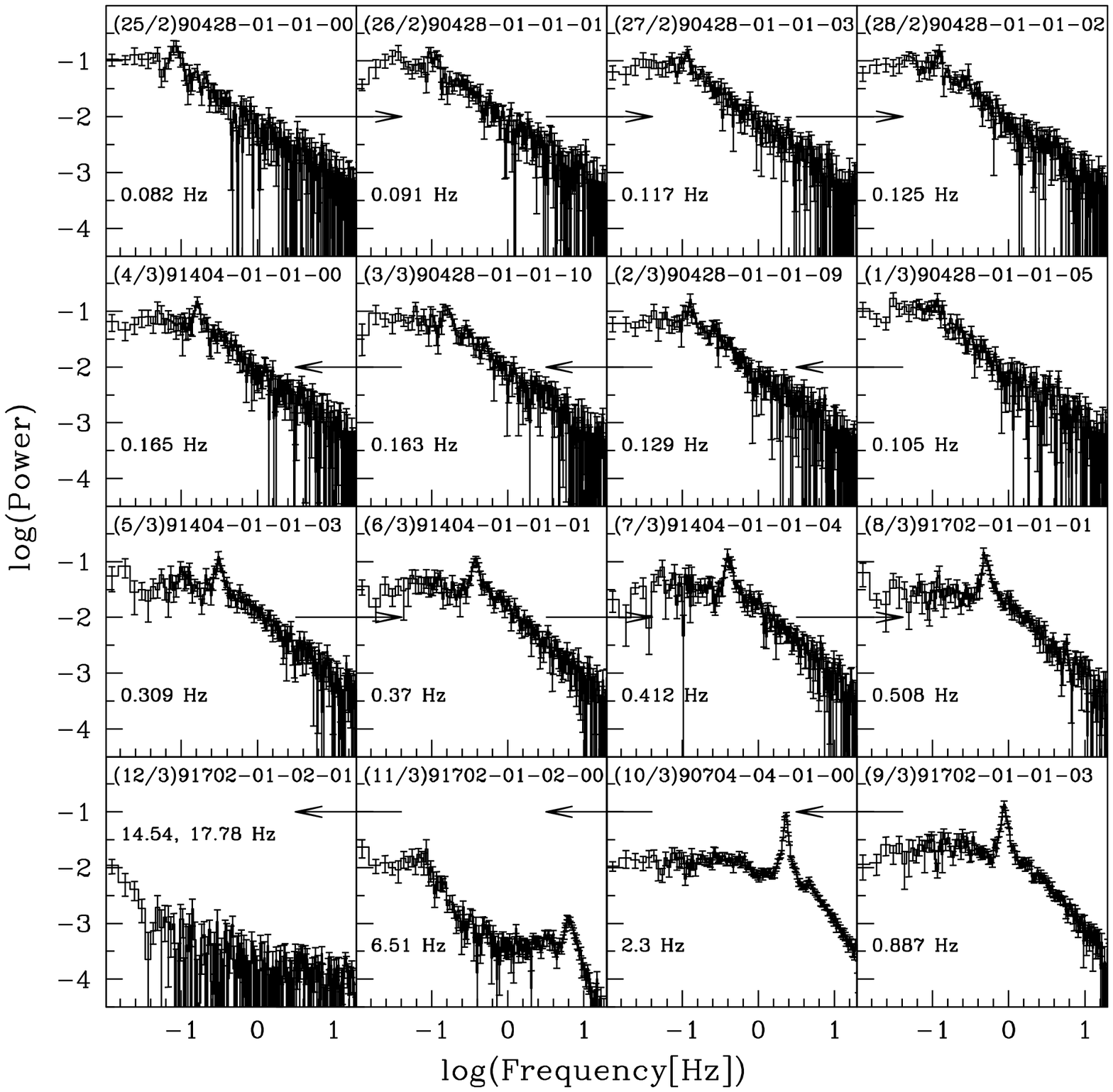}}
\vspace{0.0cm}
\noindent {\small {\bf Fig. 6:} Variation of the PDS with QPO frequencies marked in the hard state of the rising phase
from $25^{th}$ of February, 2005 to $11^{th}$ of March, 2005. The dates (dd/mm), the observation ID and the
frequency of the QPO are in the inset. Arrows indicate the direction in which the dates are increasing.}}
\end{figure}
 
\begin{figure}
\vbox{
\vskip -0.0cm
\centerline{
       \includegraphics[scale=0.6,angle=0,width=14truecm]{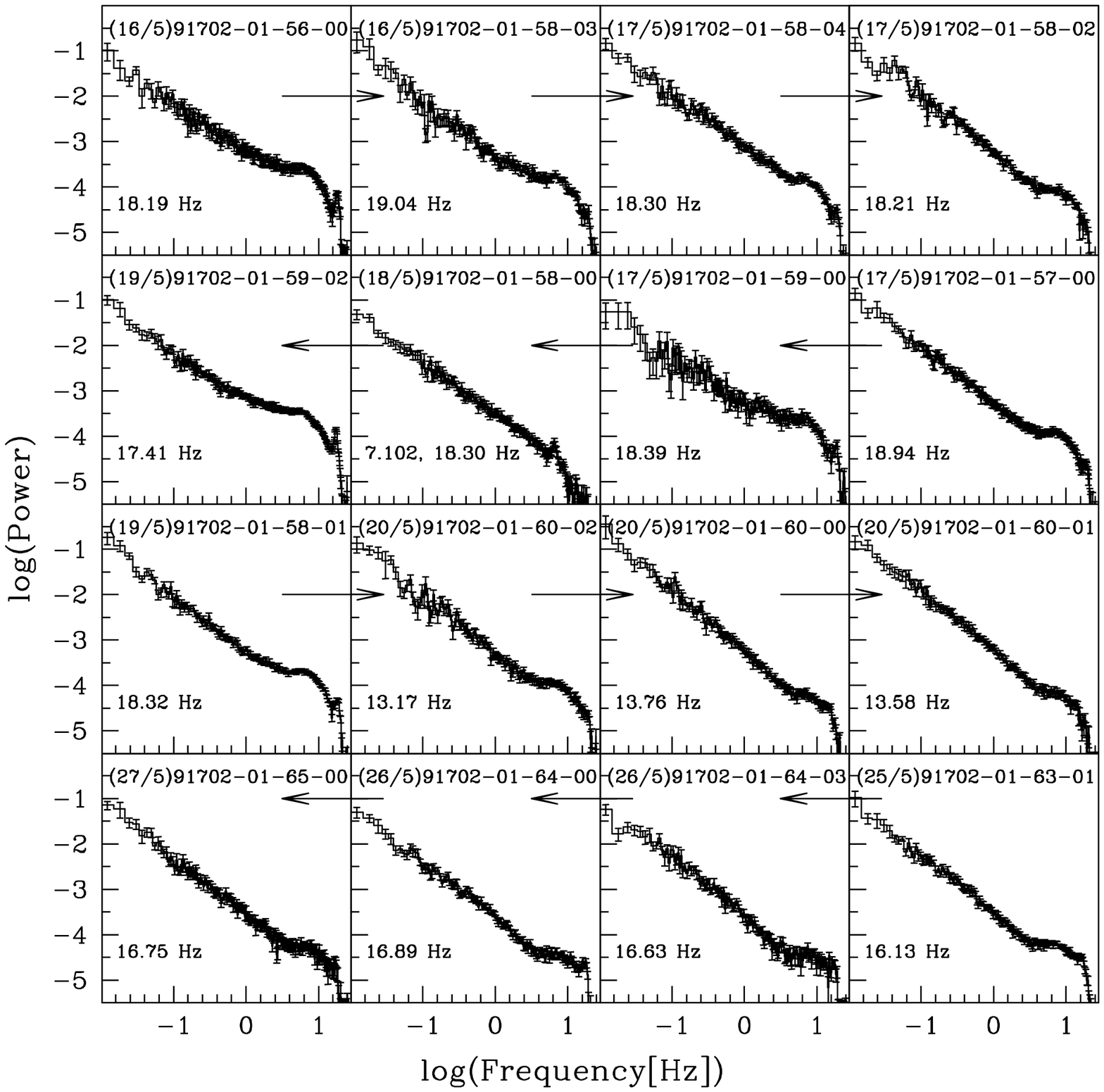}}
\vspace{0.0cm}
\noindent {\small {\bf Fig. 7:} Same as in Fig. 6, except that the data of the 
intermediate state from the 16$^{th}$ of May, 2005 to 27$^{th}$ of May, 2005 was chosen.}}
\end{figure}
 
\begin{figure}
\vbox{
\vskip -0.0cm
\centerline{
       \includegraphics[scale=0.6,angle=0,width=14truecm]{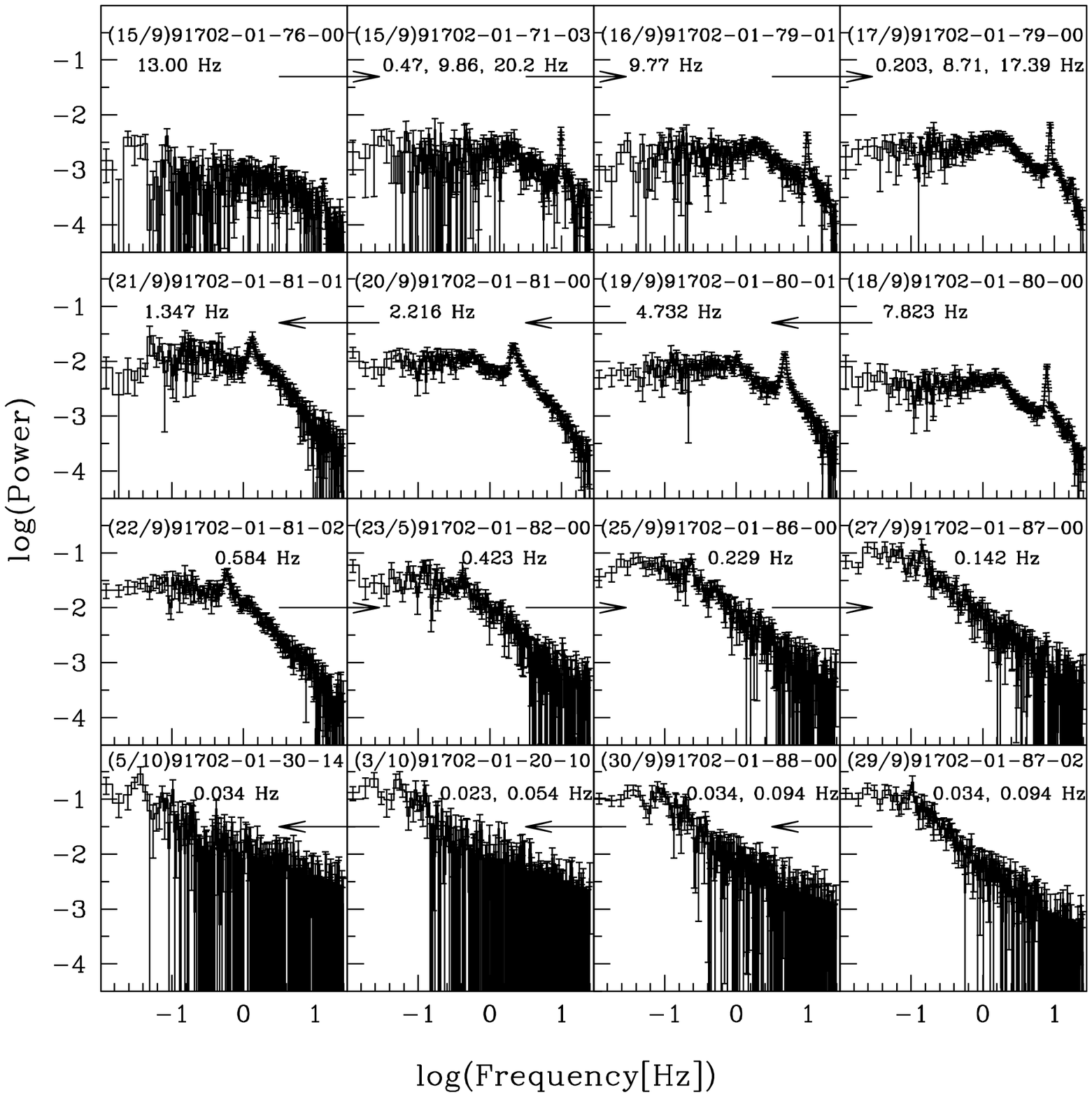}}
\vspace{0.0cm}
\noindent {\small {\bf Fig. 8:} Same as Fig. 6 except that the data of the decline phase of
the outburst from 15$^{th}$ of September, 2005 to 5$^{th}$ of October, 2005 was chosen. Frequency is
seen to be decreasing monotonically.}}
\end{figure}

\subsection {\bf\large Spectral Analysis}

For the spectral analysis we mainly used 3 - 25 keV ``{\bf Standard 2}" 
mode data from RXTE Proportional Counter Unit 2  (PCU2). In general, black hole energy spectra  (2-25 keV) 
are modeled with `diskbb' and `power-law' components, though some times best fit could be obtained when 
a Gaussian around $6.5$keV (Iron-line) was used. The results of the $150$ PCA 
observations of $123$ days are listed in Table 3. Here, we listed the components 
required for spectral fits, i.e., the disk black body Temperature $T_{in}$ in keV (Col. 2), 
normalization factor for black body fit (Col. 3), power-law photon index $\Gamma$ (Col. 4),
power-law normalization (Col. 5), disk black body flux in $3$ - $10$ keV (Col. 6), the 
power-law flux in $10$ - $25$ keV (Col. 7), the total flux in $3$ - $25$ keV (Col. 8) 
and the reduced $\chi^2$ (Col. 9).  After fitting a spectrum, 
we have used the "error" command to calculate +/- error for the
fitted parameters. All the error values are of $1\sigma$ confidence level. 
For errors bars on flux values, we use flux ``LE HE err" to calculate
+/- error \& flux for the energy range of LE and HE (in keV). 
We have put the parameter values up to 4 significant digits. 

We provide the error bar in each column. The error bars were
Fluxes are listed in units of number of photon counts/cm$^2$/sec. 
From the nature of the variation of the power-law indices and the disk black body components, 
we classified the full outburst into four spectral states: (i) Hard state from
from 25$^{th}$ of February, 2005 (MJD = 53426) to 12$^{th}$ of March, 2005 (MJD = 53441);
(ii) Soft/Very soft state from 13$^{th}$ of March, 2005 (MJD = 53442) to 15$^{th}$ of May, 2005 (MJD = 53505). 
(iii) Intermediate  state is from 16$^{th}$ of May, 2005 (MJD = 53506) to 11$^{th}$ of 
September, 2005 (MJD = 53624) and finally (iv) Hard state from the 12$^{th}$ of September, 2005 
(MJD = 53625) till 16$^{th}$ of October, 2005 (MJD = 53659). We kept the hydrogen column density 
(N$_{H}$) fixed at 7.5$\times$ 10$^{21}$ atoms cm$^{-2}$ and the systematics at $0.01$. 

Daily variations of the fitted parameters presented in Table 3 are plotted in Fig. 9  
which clearly reveals the justification of separating the full outburst in the above 
mentioned four states. The panels (a-d) are respectively the black body temperature 
$T_{in}$ in keV, the black body normalization factor, the photon index $\Gamma$ and the 
power-law normalization (plotted in the log scale along Y-axis). 
Daily variations of the total flux (panel a), black body flux 
(panel b) and the power-law flux (panel c) are shown in Fig. 10. The panel (d) shows how the 
ratio of the black body to total flux changes daily. Generally in the soft and very soft 
states the ratio is almost unity, indicating the dominance of the soft component in both of these
divisions. However, in Fig. 9, we observe a distinct difference in power-law normalization
and power-law index in these two states. The QPOs are observed only in certain days of the intermediate state.   
It may be noted that $\Gamma$ obtained right in the middle of the soft/very soft state is
unphysically high ($>4$). We believe that this is due very poor statistics (e.g., only one good PCU of RXTE was working
and photon energy was $>20$keV) rather than any unusual absorbtion at high energies. We find that other workers 
(Saito et al. 2006) also reported a high photon index for these observations. 

\begin{figure}
\vbox{
\vskip -0.0cm
\centerline{
        \includegraphics[scale=0.6,angle=0,width=14truecm]{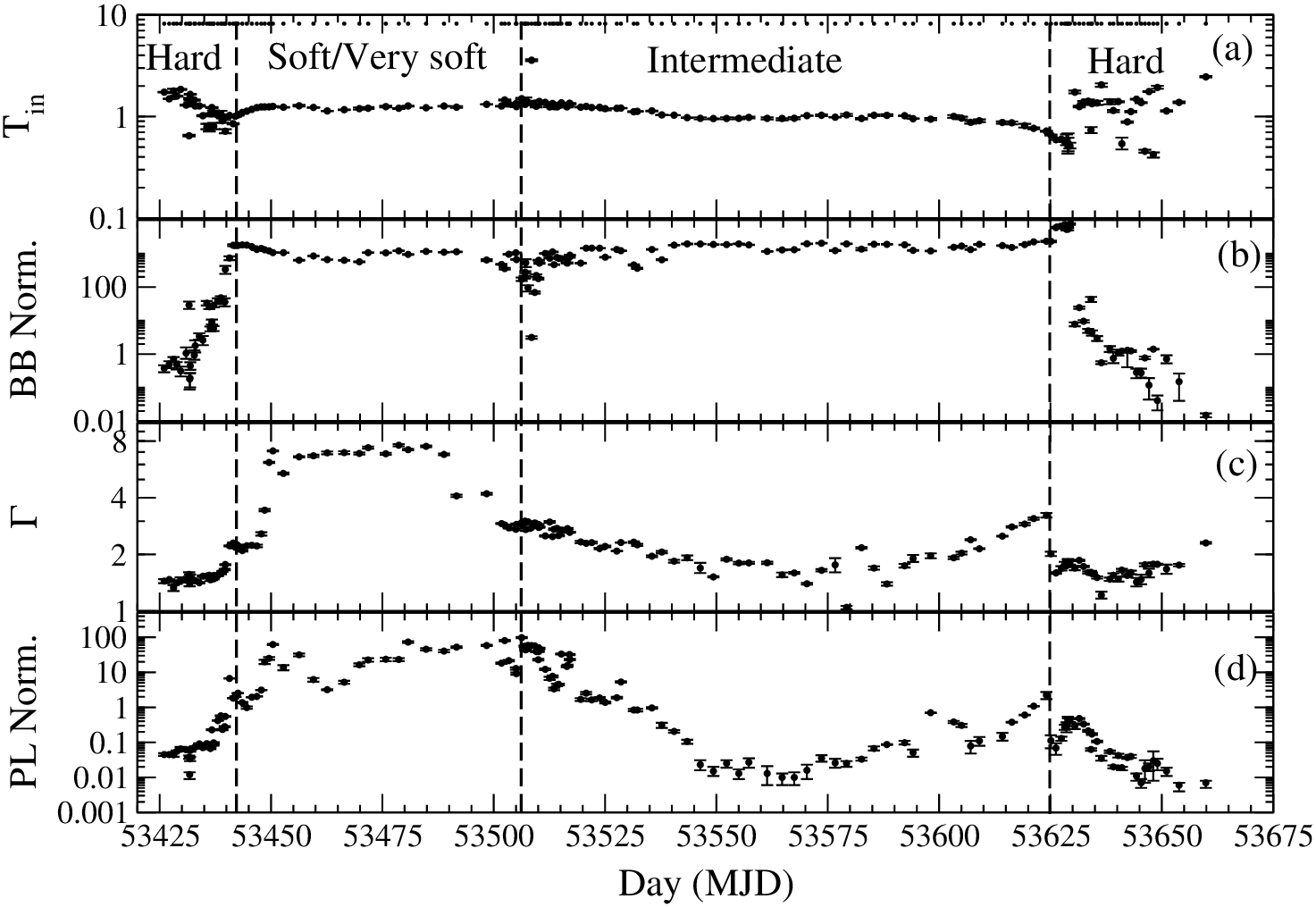}}
\vspace{2.0cm}
\noindent{\small {\bf Fig. 9:} Fitted parameters of RXTE 3 - 25 keV PCA Spectra 
plotted with time (MJD). The panels are: (a) disk black body temperature (T$_{in}$) in keV, 
(b) disk black body normalization, (c) Power-Law Photon Index ($\Gamma$) and 
(d) Power-Law normalization plotted with day. Logarithmic scale was used in the y-axis
and the error bars are at $1 \sigma$ level.}}
\end{figure}

\begin{figure}
\vbox{
\vskip 0.0cm
\centerline{
        \includegraphics[scale=0.6,angle=0,width=14truecm]{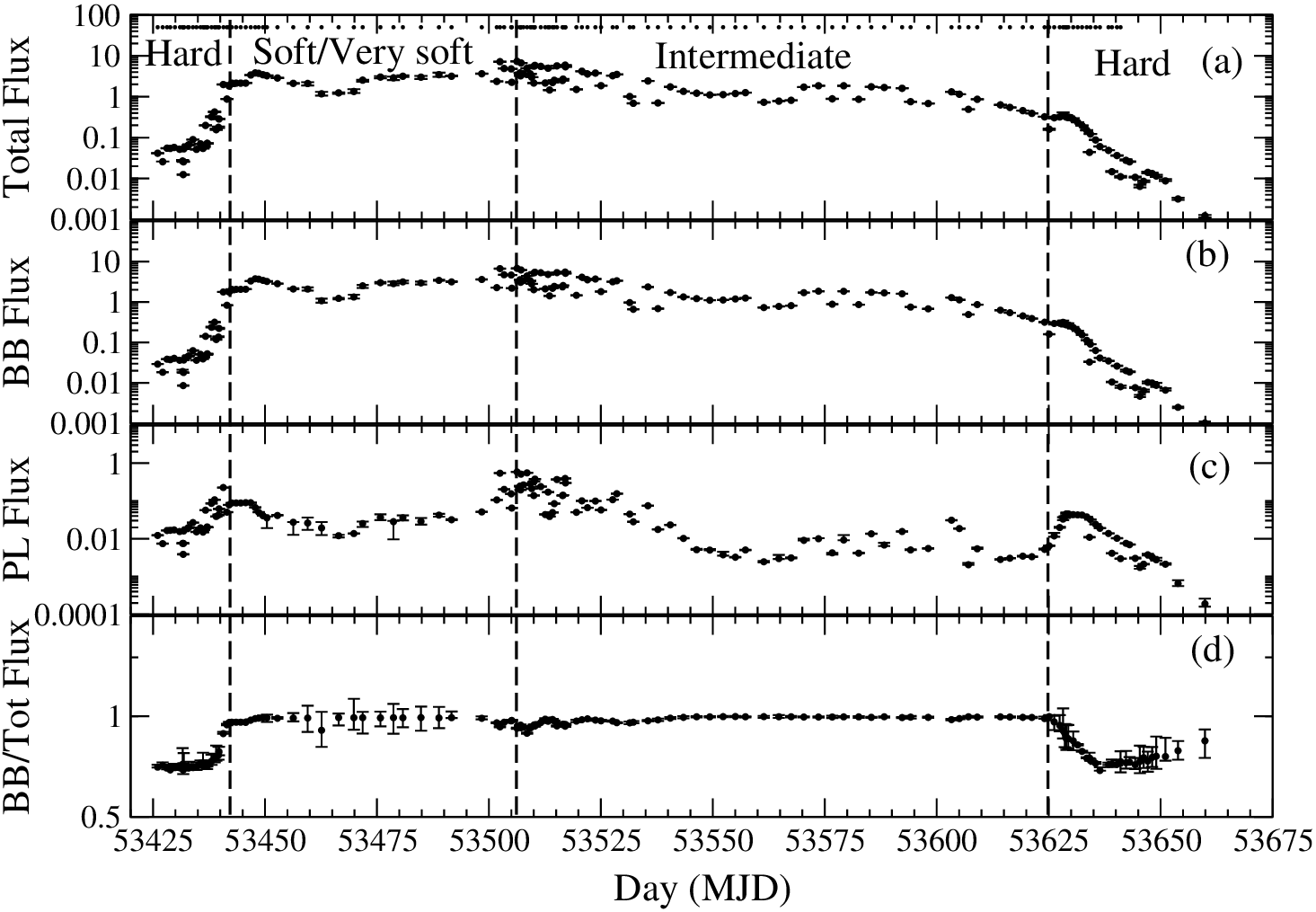}}
\vspace{2.0cm}
\noindent {\small {\bf Fig. 10:} Derived properties of the daily flux variation
are shown. The panels are: (a) 3 - 25 keV total flux, (b) 3-10 keV bolometric disk 
black body flux, (c) 10-25 keV power-law flux and (d) the ratio of the total and power-law fluxes. 
In the soft/very soft and intermediate states the total flux is dominated by the black body 
flux. Only in hard states of the rising and declining phases the ratio is less than unity. Here 
we use logarithmic scale along the y-axis.}}
\end{figure}

We have already discussed the daily variation of the spectral index and flux components. 
It is instructive to study the nature of the complete spectrum itself which we plot in Figs. 11(a-e).
In the left panels of each Figure we show the fitted spectrum with individual 
components (marked on the curves) and in the right panel we show the normalized counts/s/keV 
and the reduced $\chi^{2}_{red}$ variation. In Fig. 11(e), the component marked `Compton' comes from fitting 
with `CompST' model which represents a Compton cloud which is different from the cloud generating 
the  power-law. These components were chosen so as to get a minimum value of reduced $\chi^2$. 
To find out the requirement of extra model component to fit the data, is carried
out with the F-test task. F-test results are summarized in the Table 4. We chose the
combination of the components for which the F-test probability is lowest (see, Col. 6).

Insets show the average $\chi_{red}^2$. The Figures are drawn with data on the 10$^{th}$ March (Id: 90704-04-01-00), 
(b) $11^{th}$ March, 2005 (ID: 91702-01-02-00G), (c) $20^{th}$ March, 2005 (ID: 91702-01-08-00),
(d) $17^{th}$ May, 2005 (ID: 91702-01-57-00G), and (e) $17^{th}$ September, 2005 (ID: 91702-01-79-00).
Figure symbols have their usual meanings. What we see is that in the hard states the cooler
`Compton' component is missing, while towards the end of the `intermediate' state and the
beginning of the hard state of the decline phase this component shows up, albeit of decreasing importance.

\begin{figure}
\vskip -0.2cm
\centerline{
       \includegraphics[scale=0.6,angle=270,width=5.5truecm]{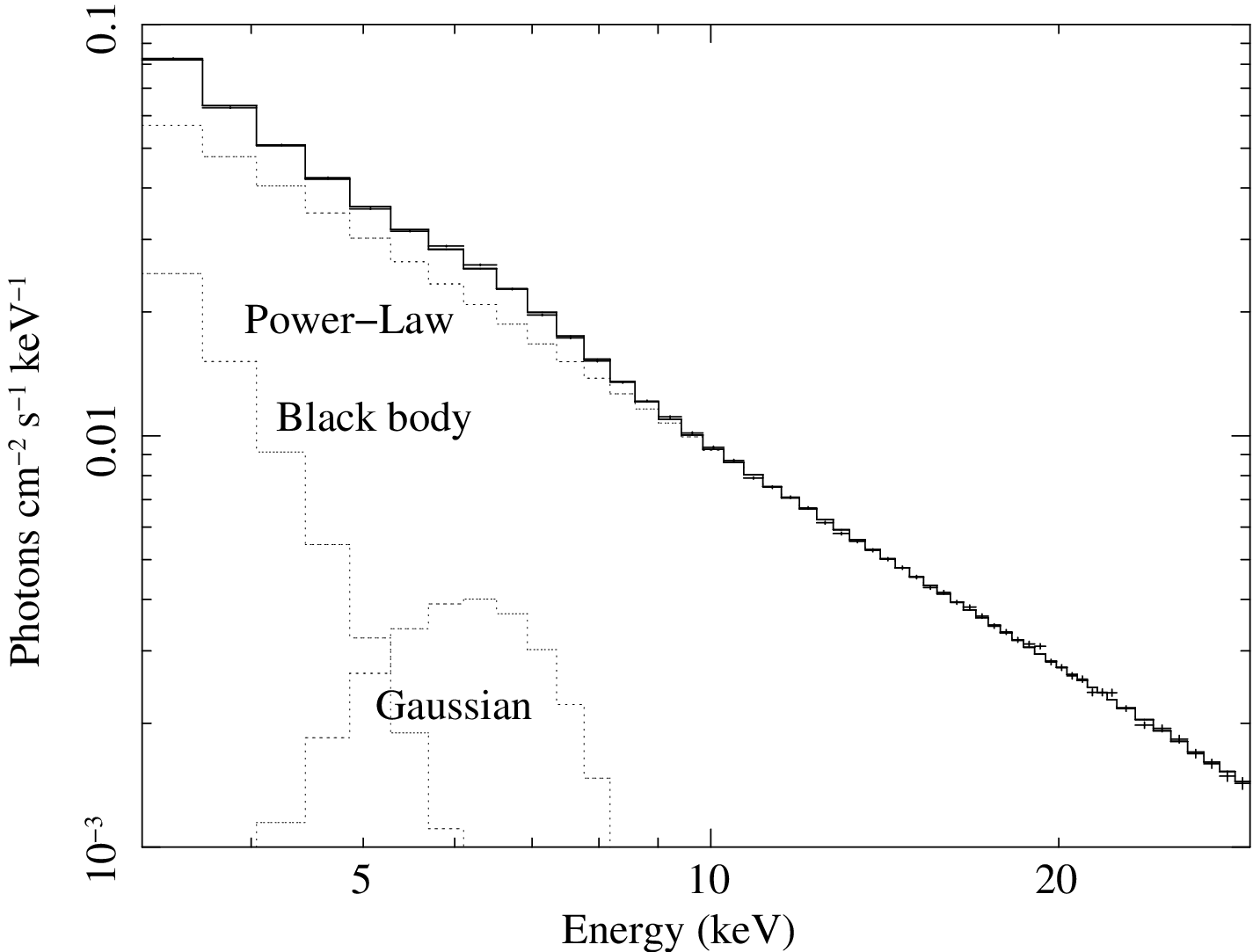}
       \includegraphics[scale=0.6,angle=270,width=5.5truecm]{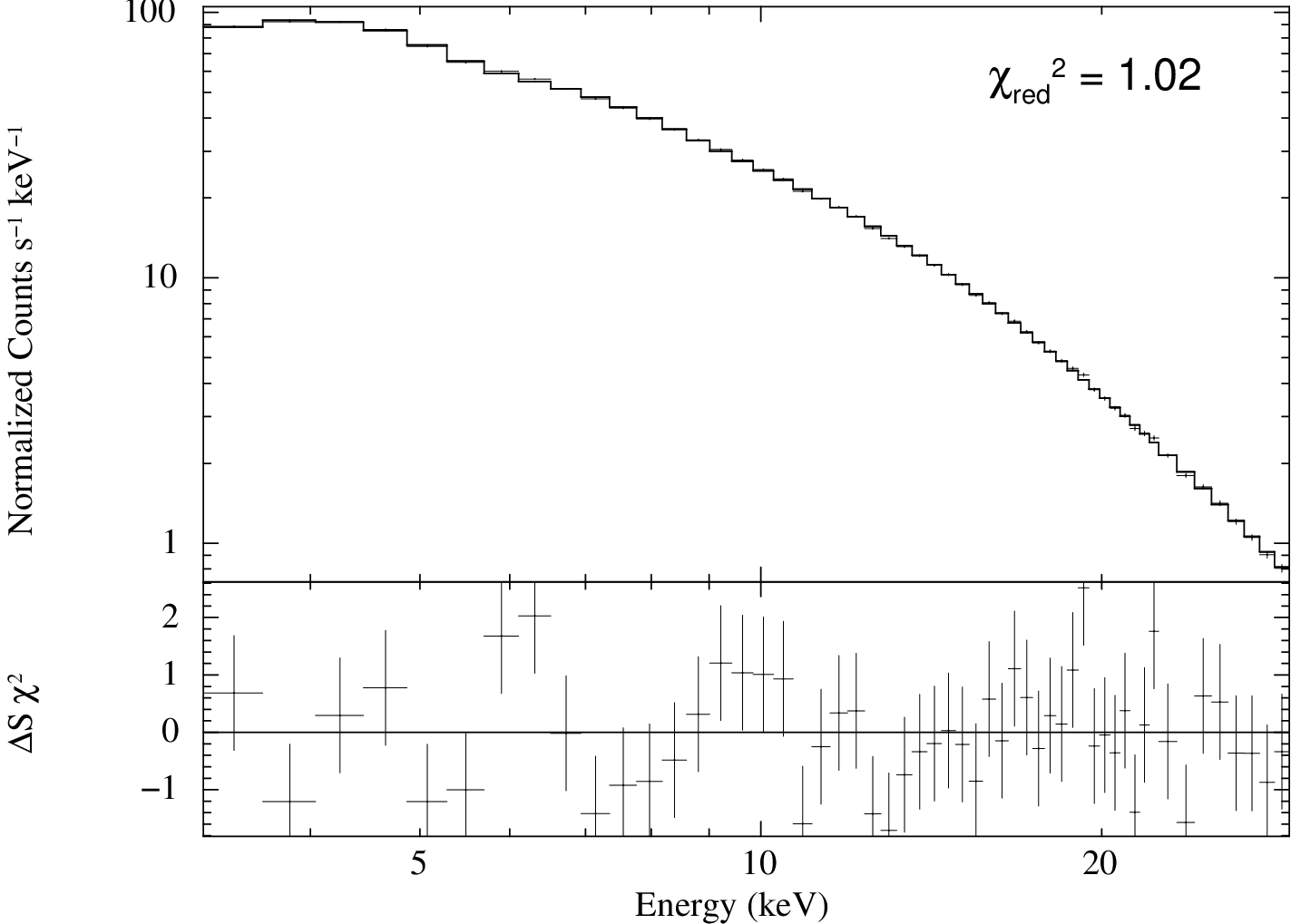}\\ }
\centerline{
       \includegraphics[scale=0.6,angle=270,width=5.5truecm]{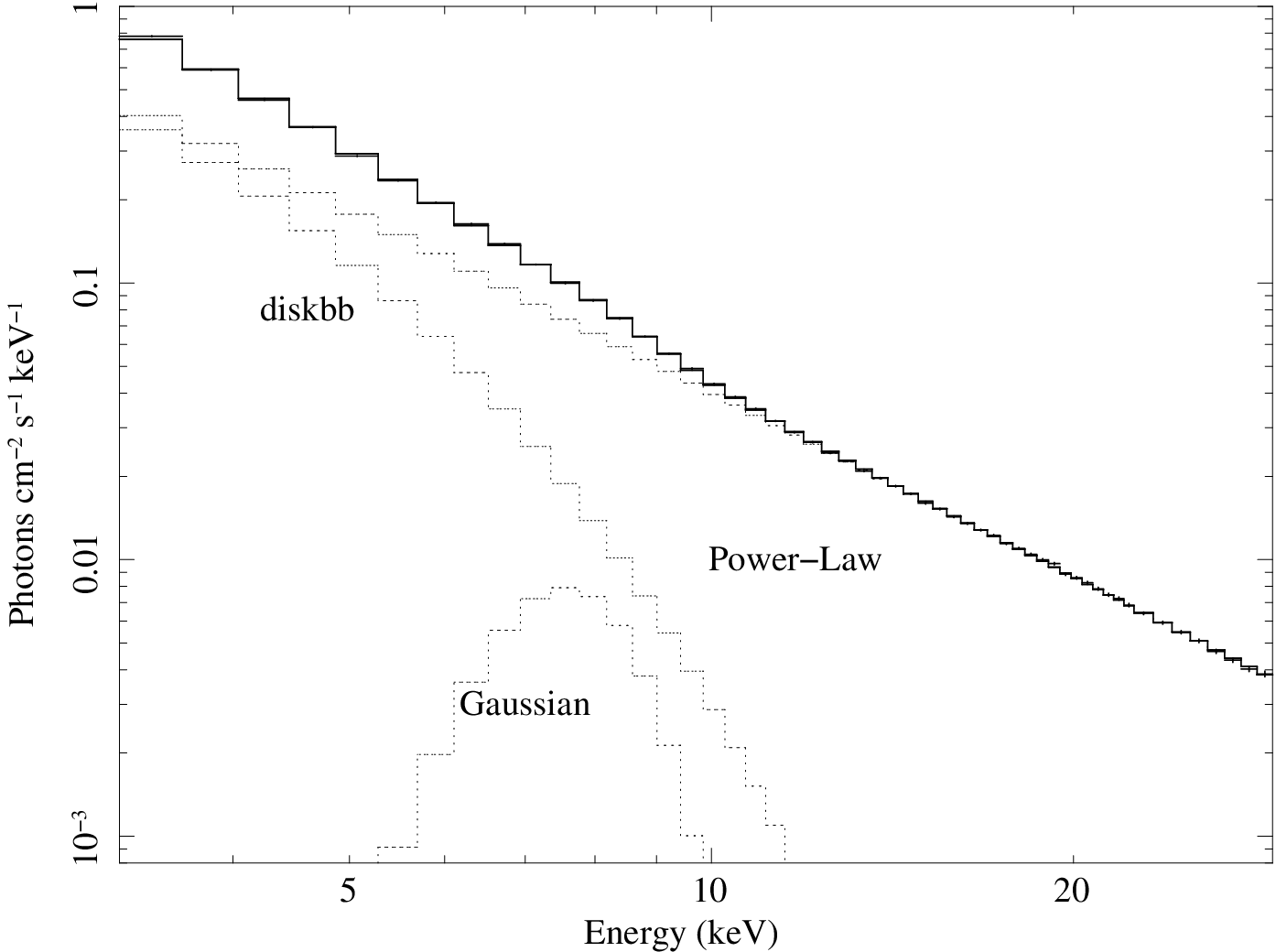}
       \includegraphics[scale=0.6,angle=270,width=5.5truecm]{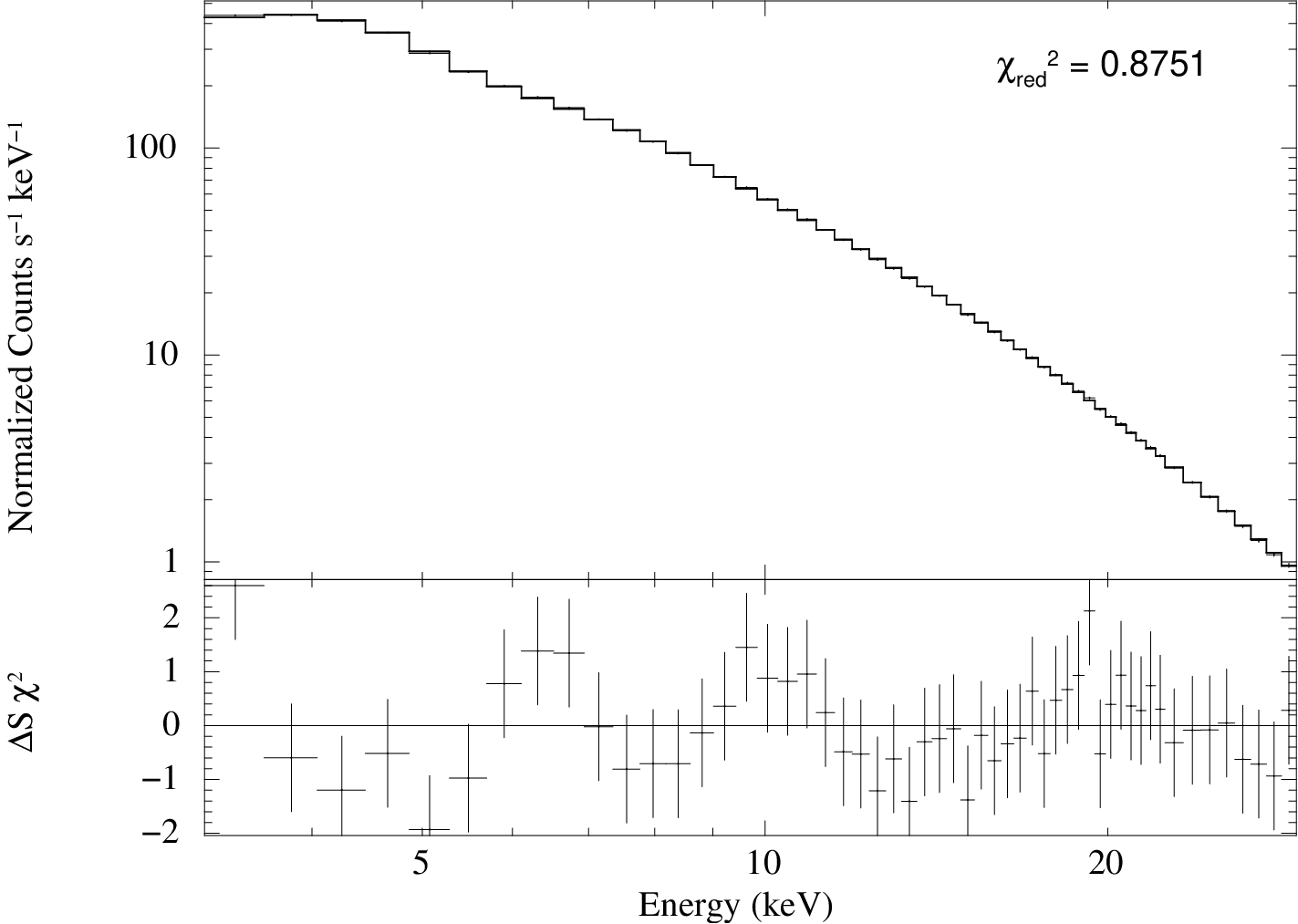}\\ }
\centerline{
       \includegraphics[scale=0.6,angle=270,width=5.5truecm]{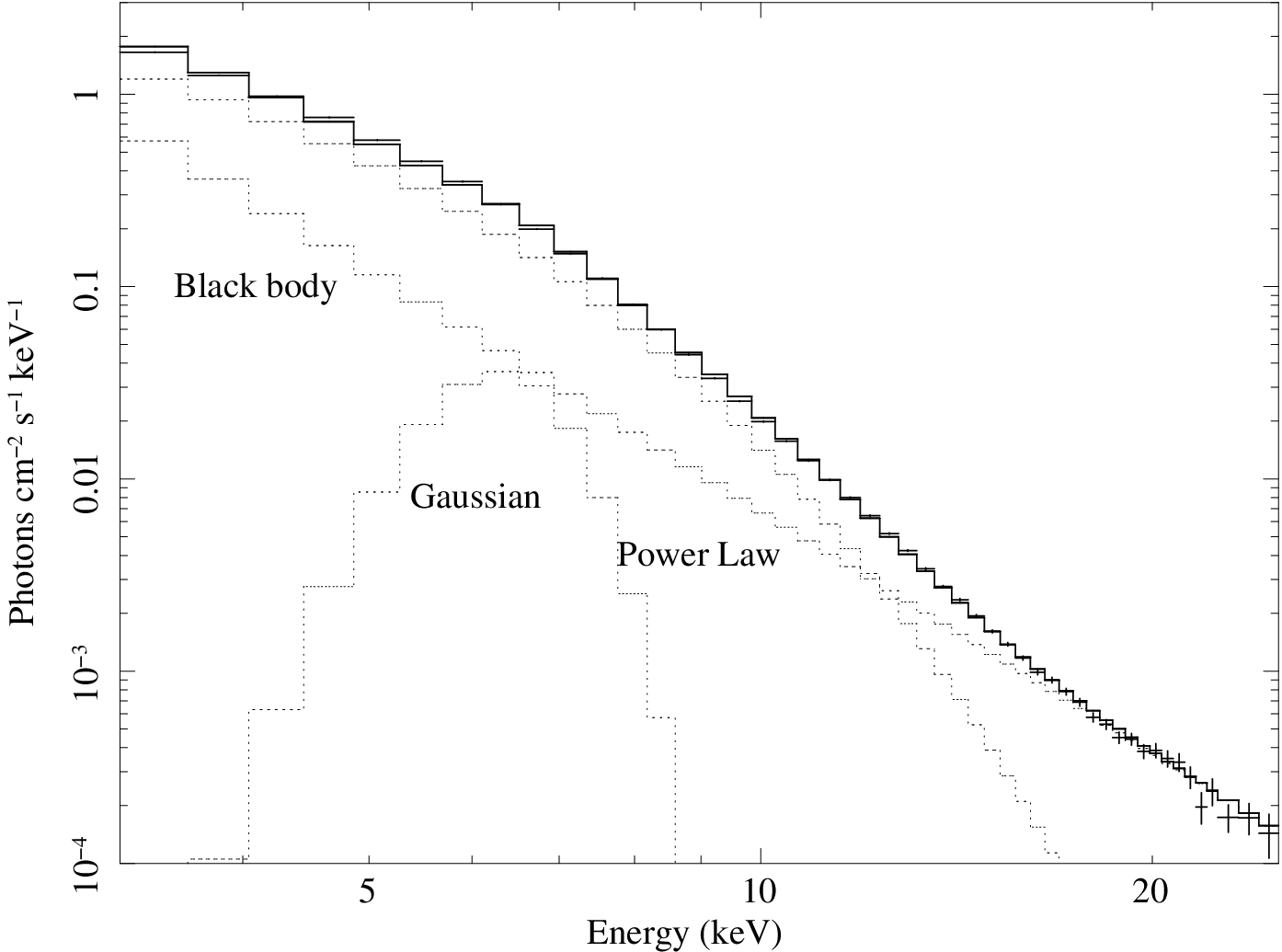}
       \includegraphics[scale=0.6,angle=270,width=5.5truecm]{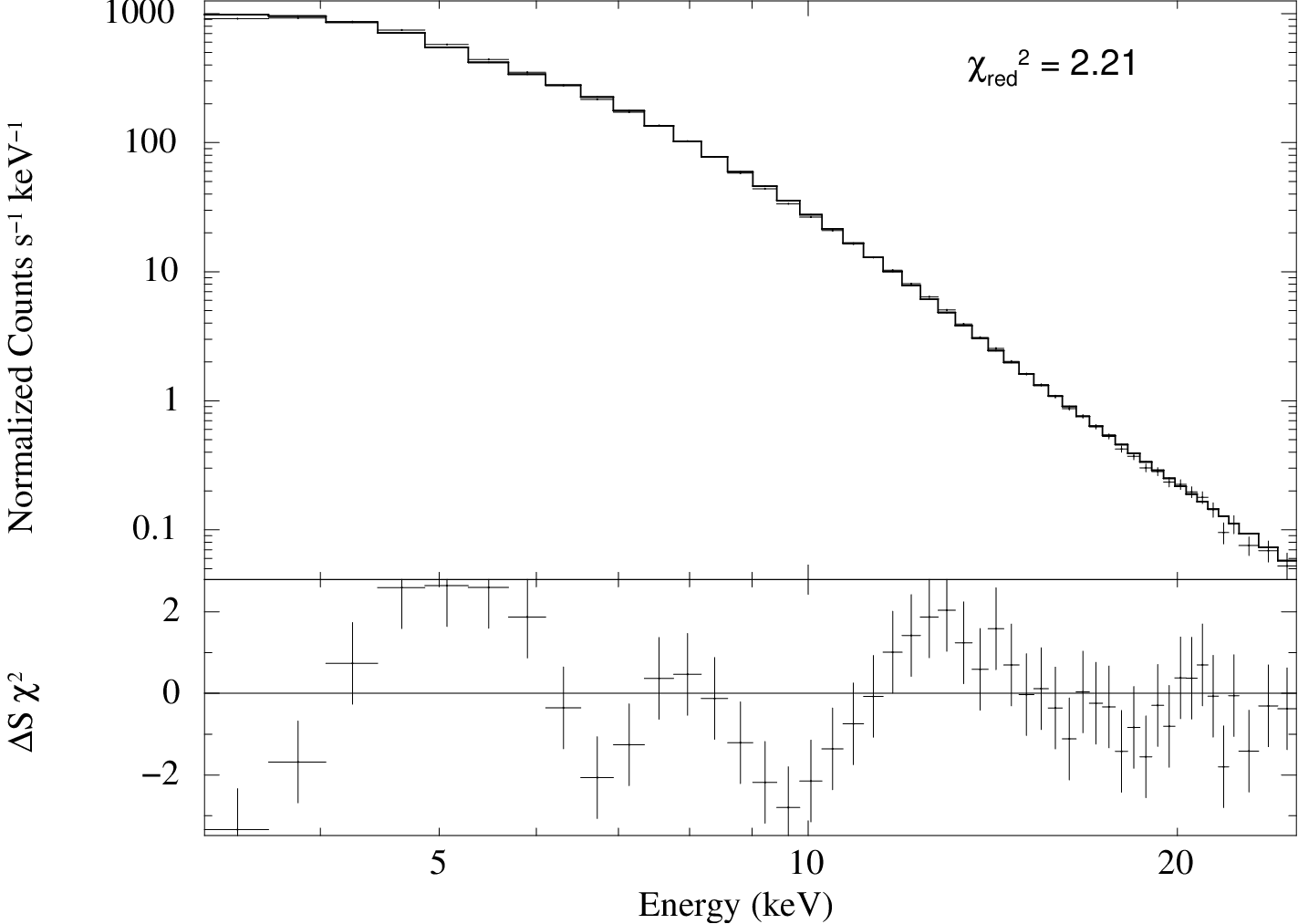}\\ }
\centerline{
       \includegraphics[scale=0.6,angle=270,width=5.5truecm]{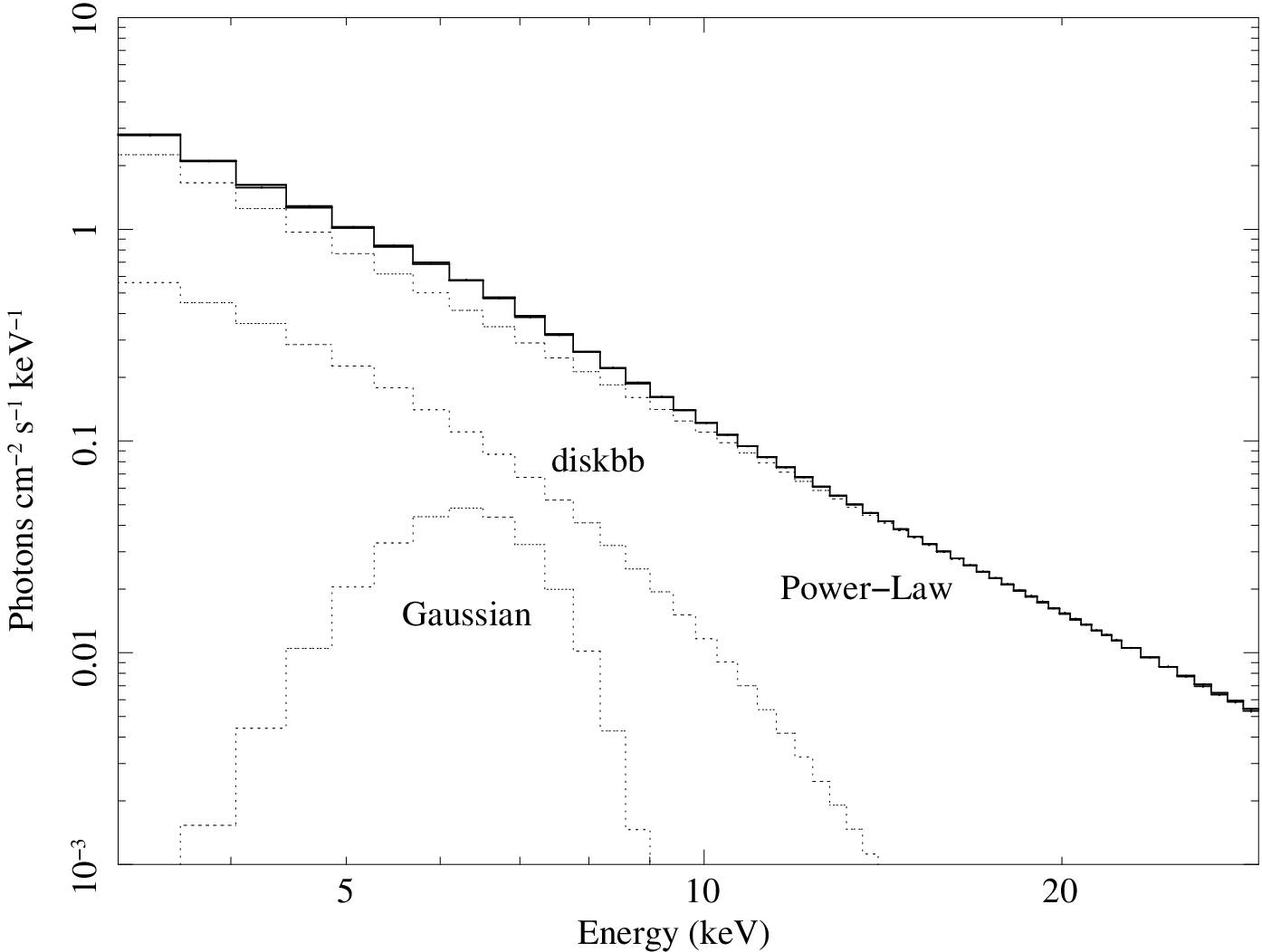}
       \includegraphics[scale=0.6,angle=270,width=5.5truecm]{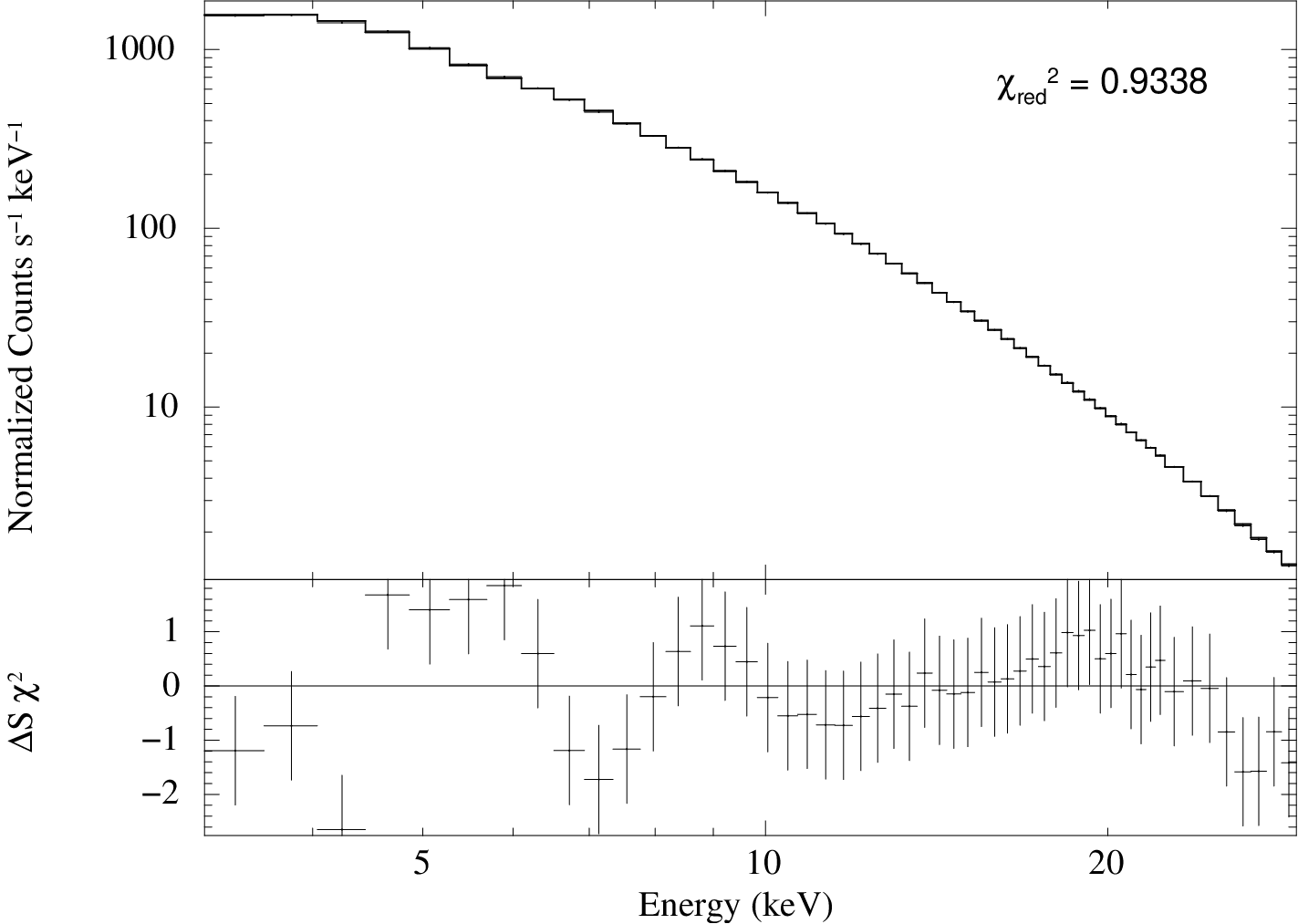}\\ }
\centerline{
       \includegraphics[scale=0.6,angle=270,width=5.5cm]{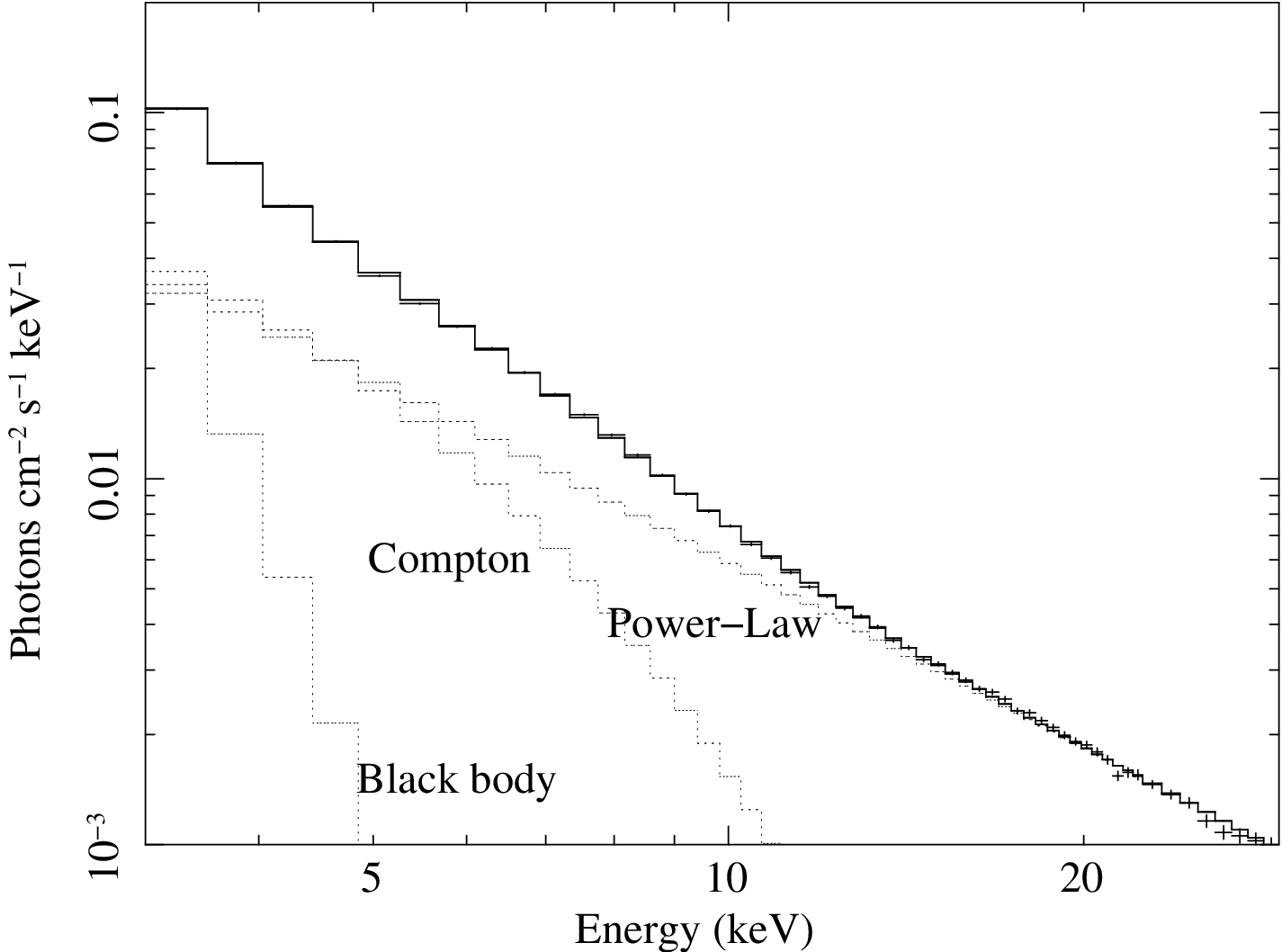}
       \includegraphics[scale=0.6,angle=270,width=5.5cm]{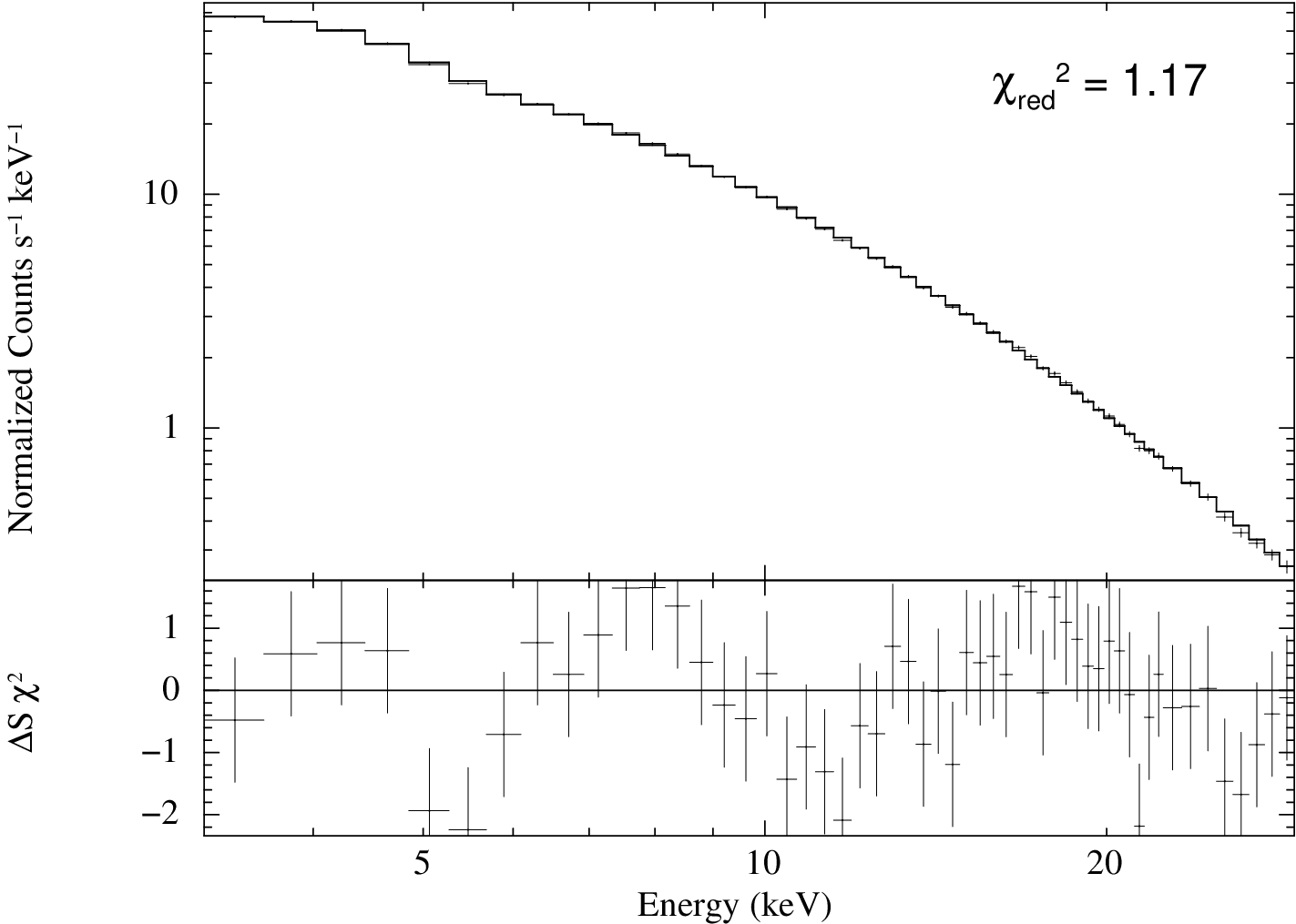}\\}

\vspace{0.0cm}
\noindent {\small {\bf Fig. 11(a-e):}  3 - 25 keV RXTE/PCA model fitted spectra 
with various components on the left panels and the fitted reduced $\chi^{2}_{red}$
on the right panels. Data used are of (a) $10^{th}$ March, 2005 (Obs ID: 90704-04-01-00),
(b) $11^{th}$ March, 2005 (ID: 91702-01-02-00G), (c) $20^{th}$ March, 2005 (ID: 91702-01-08-00),
(d) $17^{th}$ May, 2005 (ID: 91702-01-57-00G), and (e) $17^{th}$ September, 2005 
(ID: 91702-01-79-00) respectively.}

\end{figure}

\section {\bf\large Brief interpretation of the Results}

GRO 1655-40 is a typical outburst source which was observed very
regularly with one of the most successful X-ray instruments till date. The detailed
results of RXTE that we presented reveal several very important aspects of the 
nature of the transient accretion process around a black hole. From the light curves,
hardness/softness diagrams, spectral slopes and most importantly the variation of the QPO 
frequency, one can come up with very comprehensive picture of what might be 
happening when such an outburst takes place. 

First, we concentrate on the rising and decline phases of the outburst. If we make 
the most natural assumption that rushing in of matter towards a black hole is 
the cause of the outburst, then during rising phase the matter is increasing 
while in the decline phase the matter is evacuated with little fresh supply. The 
formation of strong QPOs and the smooth variation of QPO frequencies during 
the outburst (Chakrabarti et al. 2005, 2008) indicates that the cause of 
QPO is identical each day and is related to the a dynamical property of the
infalling matter. While a popular model for low frequency QPO assumes the motion
of a perturbation or blob at the inner edge of a Keplerian disk (e.g., Trudolyubov et al. 1999)
it is difficult to imagine how a perturbation would sustain itself 
against shear and dissipation for more than a few orbits, let alone more than 
two weeks which we observe here. Because of this, we prefer the oscillating 
shock solution inside a sub-Keplerian disk which has been demonstrated
to have a stable oscillation for many dynamical time scale (Molteni, Sponholz \& Chakrabarti 1996; 
Ryu, Chakrabarti \& Molteni, 1997; Chakrabarti \& Manickam, 2000; Chakrabarti, Acharyya \& Molteni, 2004). 
It is easy to verify that the QPO frequencies (which are inverses of the infall times from the 
post-shock flow to the black hole) in the infalling phase are simply related, as though
the shock itself is drifting towards the black hole at a slow pace of $\sim 20m/s$ (Chakrabarti et al. 2005).
In the decline phase, in the same way, the shock was found to recede, at first very slowly 
(as though there was still some significant infalling matter) for about three days, and then 
at an almost constant acceleration (Chakrabarti et al., 2008).

During the rising phase of more than two weeks, the disk got sufficient time to transport 
angular momentum and a dominant Keplerian disk is formed which made the flow soft or very
soft. The rapid rise of the black body flux after the QPO disappears and almost total absence 
of the hard photons testify to the rushing in of the Keplerian disk towards the inner 
edge (Chakrabarti \& Titarchuk, 1995; Ebisawa et al. 1996).
If we take the two component advective flow (TCAF) model one step further and actually fit the
spectra of a few days spreaded during the outburst we observe, using the same procedure
that was followed in Chakrabarti \& Mandal (2006), we can obtain the accretions rates of 
matter in the Keplerian disk and the sub-Keplerian halo. Table 5 gives the rates in 
units of Eddington rate on various days. It is clear that the Keplerian disk rate
steady increases from the beginning while the halo rate changes in a shorter time scale. At the beginning,
the halo rate was higher than the disk rate, but in the rest of the time, until the
very end the disk rate always dominates. In the soft and the very soft states, 
the disk rate required to fit the spectra can be high reaching to about two Eddington rates.
The hardness/softness diagrams also give an idea of how the accretion rates in the 
Keplerian and sub-Keplerian components could be changed on a daily basis.
After the very soft state is passed, the viscous processes became weaker 
and inflowing matter which continues to accrete sporadically becomes dominant. The count rate 
rapidly fell from tens of thousands to a few hundreds. QPOs started appearing only 
sporadically in this state. The general trend of the declining inflow rate
together with the lowering of viscosity ensured the sucking in of the Keplerian matter. In the hard state 
of the declining phase, the sub-Keplerian component became comparable to the Keplerian
rate giving rise to a strong power-law flux and QPOs.

In the literature, there are reports of other sources which exhibited similar outbursts. 
For XTE J1550-564, a similar interpretation with TCAF and shock waves was found to be very successful
(Soria et al. 2001, Wu et al. 2002, Chakrabarti, Datta \& Pal 2009). The spectral state transitions
in outburst sources appear to be fundamentally different from those in a 
persistent source (such as Cyg X-1). In the latter case, the total flux could be 
almost constant even during the state transitions (Zhang et al. 1996) where the Keplerian
and the sub-Keplerian rates could be redistributed during the state transition and as
a result the total flux could be almost constant. In outburst sources, on the other hand,
the approaching and receding Keplerian component in the rising and decline phases causes
the net flux to rise and fall during the hard to soft and intermediate to hard state
respectively.  

\section {\bf\large Concluding Remarks}

In this paper, we presented a comprehensive analysis of the entire 2005 outburst of GRO J1655-40.
The results clearly indicated that the spectral state passed from the hard state 
in the rising phase with a poor Keplerian component to 
the soft/very soft and intermediate states dominated by Keplerian
disks and finally to the hard state at the decline phase. 
It is often believed that low frequency QPOs may be generated by perturbations
at the inner edge of a Keplerian disk, either by orbiting  `blobs' or more 
probably oscillating shocks. Our observation of the smooth
variation of the QPO frequency during the rising and the decline phases 
indicates that whatever be the reason for QPO, it has to survive for weeks.
It is difficult to imagine that disk perturbation with blobs
be sustained for such a long time without being sheared 
and dissipated. On the other hand, shock oscillations have been shown to 
survive for a long period and the PDS calculated from observations
also resembles the observed PDS. QPO frequencies at the rising and 
decline phases have been shown separately (Chakrabarti et al. 2005, 2008, 2009) 
to be simply related as though an oscillating shock is drifting in at the rising phase and 
drifting out at the decline phase. We therefore favour the shock oscillation solution of the 
low and intermediate frequency QPOs. Recently (Titarchuk, Shaposnikov \& Arefiev, 2007) have pointed 
out that the presence of bumps in the PDS of Cyg X-1 is the 
signature of a sub-Keplerian flow in the accretion disk. 
We have also observed similar bumps on days which have significant halo component
(see, Fig. 4c and Table 5). Thus we believe that the general picture
which emerged out of our analysis is consistent with a two component advective flow
as has been pointed out by many authors in the context of several black holes
(Smith, Heindl, Markwardt \& Swank, 2001; Smith, Heindl \& Swank, 2002; Smith, Dawson
\& Swank, 2007). The TCAF model is further supported by a clear indication of a jump 
in total flux at the state transitions (hard to soft and intermediate to hard) 
in this source. This we believe is due to the approaching and receding Keplerian 
component in the rising and declining phases.  This is in contrast with persistence 
X-ray sources, such as Cyg X-1, where the total flux remains constant.

\newpage
\begin{table}
\vskip -0.5cm
\scriptsize
\centering
\centerline {Table 1}
\centerline {PCA Count rates, Hardness ratios and QPOs}
\vskip 0.2cm
\begin{tabular}{lccccccccl}
\hline
\hline
  &   &   &    &  PCA  &  PCA   & PCA  &    &    &     \\
Obs.&UT Date&MJD Start&MJD Stop&2-3.5 keV&3.5-10.5 keV&10.5-60 keV&
$HR5^{a}$&$HR6^{b}$&QPO Frequency\\
 (1)  &(2)  &(3) & (4) & (5) & (6) & (7)  & (8)  &  (9)  & (10)  \\ 
\hline

1 &25/02/05&53426.0400&53426.0806&  4&  30& 29& 7.341& 1.026&0.082\\
2 &26/02/05&53427.1546&53427.1939&  5&  37& 34& 7.234& 1.088&0.034, 0.106\\
3 &27/02/05&53428.1380&53428.1786&  5&  39& 35& 7.350& 1.118&0.116\\
4 &27/02/05&53428.8594&53428.8994&  5&  40& 35& 7.381& 1.161&0.051, 0.122\\
5 &28/02/05&53429.9080&53429.9480&  5&  40& 35& 7.341& 1.146&0.051, 0.121\\
\hline
\end{tabular}
\noindent{
\leftline {$^{a}$ Ratio of 3.5-10.5 keV and 2-3.5 keV PCA count rates.}
\leftline {$^{b}$ Ratio of 3.5-10.5 keV and 10.5-60 keV PCA count rates.}}
\end{table}

\newpage
\begin{table}
\vskip -0.5cm
\scriptsize
\centering
\centerline {Table 1 (Cont'd)}
\centerline {PCA Count rates, Hardness ratios and QPOs}
\vskip 0.2cm
\begin{tabular}{lccccccccl}
\hline
\hline
     &   &   &   &  PCA    &    PCA     &    PCA    &      &      &      \\
Obs.&UT Date&MJD Start&MJD Stop&2-3.5 keV&3.5-10.5 keV&10.5-60 keV&
$HR5^{a}$&$HR6^{b}$&QPO Frequency\\
 (1)  &  (2)  &  (3)  &(4)  & (5) &  (6)  & (7)  &  (8)  &  (9)  & (10) \\ 
\hline
 
51&08/05/05&53498.4152&53498.6526 & 694&2615& 70& 3.768&37.357&...\\
52&11/05/05&53501.7698&53501.7991 & 866&3507&250& 4.049&14.028&...\\
53&12/05/05&53502.4883&53502.5206 &1266&5220&517& 4.123&10.096&...\\
54&13/05/05&53503.3948&53503.4491 & 914&3602&219& 3.940&16.447&...\\
55&14/05/05&53504.9885&53505.0170 & 884&3542&171& 4.006&20.713&...\\
\hline
\end{tabular}
\noindent{
\leftline {$^{a}$ Ratio of 3.5-10.5 keV and 2-3.5 keV PCA count rates.}
\leftline {$^{b}$ Ratio of 3.5-10.5 keV and 10.5-60 keV PCA count rates.}}
\end{table}

\newpage
\begin{table}
\vskip -0.5cm
\scriptsize
\centering
\centerline {Table 1 (Cont'd)}
\centerline {PCA Count rates, Hardness ratios and QPOs}
\vskip 0.2cm
\begin{tabular}{lccccccccl}
\hline
\hline
   &   &  &   &  PCA    &    PCA  &    PCA    &      &      &    \\
Obs.&UT Date&MJD Start&MJD Stop&2-3.5 keV&3.5-10.5 keV&10.5-60 keV&
$HR5^{a}$&$HR6^{b}$&QPO Frequency\\
 (1) &  (2)& (3) & (4) & (5) &  (6) & (7)  &  (8)  &  (9)  & (10) \\ 
\hline

101&19/07/05&53570.3257&53570.3546&412&1160& 25&2.815&46.400&...\\
102&22/07/05&53573.6024&53573.8011&439&1255& 26&2.858&48.269&...\\
103&25/07/05&53576.6861&53576.7248&428&1230& 27&2.873&45.555&...\\
104&28/07/05&53579.2404&53579.2644&441&1261& 26&2.859&48.500&...\\
105&31/07/05&53582.5831&53582.6220&425&1198& 26&2.818&46.076&...\\
\hline
\end{tabular}
\noindent{
\leftline {$^{a}$ Ratio of 3.5-10.5 keV and 2-3.5 keV PCA count rates.}
\leftline {$^{b}$ Ratio of 3.5-10.5 keV and 10.5-60 keV PCA count rates.}}
\end{table}

\begin{table}
\scriptsize
\centering
\centerline {Table 2}
\centerline {Observed QPO fitted parameters}
\vskip 0.2cm
\begin{tabular}{lccccc}
\hline
\hline
 & & & & & RMS \\
Obs.&UT Date&$\nu$&$\Delta\nu$&Q&Amplitude\\
    (1) & (2) &  (3)  &  (4)  &  (5)  &  (6)  \\
\hline

1 &25/02/05&$0.082_{-0.002}^{+0.007}$&$0.018_{-0.008}^{+0.013}$&4.432&6.410\\
2 &26/02/05&$0.034_{-0.011}^{+0.011}$&$0.010_{-0.002}^{+0.003}$&3.269&5.021\\
2 &26/02/05&$0.106_{-0.025}^{+0.010}$&$0.036_{-0.009}^{+0.005}$&2.961&6.852\\
3 &27/02/05&$0.116_{-0.007}^{+0.006}$&$0.017_{-0.008}^{+0.005}$&6.824&4.908\\
4 &27/02/05&$0.051_{-0.008}^{+0.012}$&$0.029_{-0.007}^{+0.006}$&1.735&6.048\\
\hline
\end{tabular}
\noindent{}
\end{table}

\begin{table}
\scriptsize
\centering
\centerline {Table 2 (Cont'd)}
\centerline {Observed QPO fitted parameters}
\vskip 0.2cm
\begin{tabular}{lccccc}
\hline
\hline
 & & & & & RMS \\
Obs.&UT Date&$\nu$&$\Delta\nu$&Q&Amplitude\\
    (1) & (2) &  (3)  &  (4)  &  (5)  &  (6)  \\
\hline

57&16/05/05&$18.19_{-0.280}^{+0.260}$&$0.922_{-0.303}^{+0.236}$&19.735&0.825\\
58&16/05/05&$19.04_{-0.200}^{+0.180}$&$1.248_{-0.216}^{+0.144}$&15.256&0.642\\
59&17/05/05&$18.30_{-0.250}^{+0.180}$&$0.573_{-0.100}^{+0.110}$&31.954&0.502\\
60&17/05/05&$18.21_{-0.280}^{+0.190}$&$0.701_{-0.019}^{+0.022}$&25.973&0.420\\
61&17/05/05&$18.94_{-0.120}^{+0.120}$&$0.245_{-0.002}^{+0.002}$&77.212&0.317\\
\hline
\end{tabular}
\noindent{}
\end{table}
\begin{table}
\vskip -0.5cm
\scriptsize
\centering
\centerline {Table 3}
\centerline {PCA Spectral fitted parameters}
\vskip 0.2cm
\begin{tabular}{lccccccccc}
\hline
\hline
Obs.&diskbb&diskbb&Power-Law&Power-Law&Black Body&Power-Law&Total&$\chi^2_{red.}$\\
    &T$_{in}(keV)$&Norm.&Index($\Gamma$)&Norm.&Flux$^{d}$&Flux$^{e}$&Flux$^{f}$&\\
    (1)    &  (2)  & (3)  & (4)&   (5)   &  (6)    &    (7)      &  (8)     &  (9)   \\ 
\hline

1$^{a}$ &$1.738_{-0.012}^{+0.012}$&$ 0.379_{-0.085}^{+0.092}$&$1.442_{-0.035}^{+0.036}$&$0.045_{-0.004}^{+0.004}$&$0.0292_{-0.0004}^{+0.0007}$&$0.0123_{-0.0001}^{+0.0001}$&$0.0415_{-0.0004}^{+0.0004}$&1.133\\
2$^{a}$ &$1.494_{-0.011}^{+0.011}$&$ 0.495_{-0.114}^{+0.112}$&$1.469_{-0.003}^{+0.003}$&$0.045_{-0.003}^{+0.004}$&$0.0184_{-0.0003}^{+0.0006}$&$0.0074_{-0.0001}^{+0.0002}$&$0.0258_{-0.0004}^{+0.0002}$&1.051\\
3$^{a}$ &$1.785_{-0.089}^{+0.088}$&$ 0.703_{-0.105}^{+0.104}$&$1.326_{-0.046}^{+0.048}$&$0.044_{-0.005}^{+0.005}$&$0.0386_{-0.0004}^{+0.0002}$&$0.0166_{-0.0001}^{+0.0002}$&$0.0551_{-0.0003}^{+0.0004}$&1.026\\     
4$^{b}$ &$1.588_{-0.011}^{+0.018}$&$ 0.482_{-0.094}^{+0.134}$&$1.411_{-0.018}^{+0.019}$&$0.056_{-0.003}^{+0.002}$&$0.0375_{-0.0012}^{+0.0013}$&$0.0168_{-0.0001}^{+0.0001}$&$0.0543_{-0.0010}^{+0.0001}$&0.601\\
5$^{b}$ &$1.848_{-0.019}^{+0.018}$&$ 0.314_{-0.109}^{+0.098}$&$1.460_{-0.049}^{+0.044}$&$0.065_{-0.009}^{+0.007}$&$0.0406_{-0.0006}^{+0.0004}$&$0.0171_{-0.0003}^{+0.0004}$&$0.0577_{-0.0012}^{+0.0002}$&1.071\\
\hline
\end{tabular}
\noindent{
\leftline {   Fitted with $^{a}$ diskbb + power law, 
$^{b}$ diskbb + Gaussian + power law and $^{c}$ diskbb + compST + power law} 
\leftline {   $^{d}$ PCA 3-10 keV,  $^{e}$ PCA 10-25 keV \&   
$^{f}$ PCA 3-25 keV Model Fitted Photon Fluxes in number of photons$/cm^2/s$.}}
\end{table}

\begin{table}
\vskip -0.5cm
\scriptsize
\centering
\centerline {Table 3 (Cont'd)}
\centerline {PCA Spectral fitted parameters}
\vskip 0.2cm
\begin{tabular}{lccccccccc}
\hline
\hline
Obs.&diskbb&diskbb&Power-Law&Power-Law&Black Body&Power-Law&Total&$\chi^2_{red.}$\\
    &T$_{in}(keV)$&Norm.&Index($\Gamma$)&Norm.&Flux$^{d}$&Flux$^{e}$&Flux$^{f}$&\\
    (1)    &  (2)  & (3)  & (4)&   (5)   &  (6)    &    (7)      &  (8)     &  (9)   \\ 
\hline
 
51$^{b}$& $1.319_{-0.007}^{+0.007}$&$641.910_{-40.744}^{+39.780}$&$4.199_{-0.045}^{+0.043}$&$58.50_{-2.283}^{+2.006}$&$3.6106_{-0.0506}^{+0.0284}$&$0.0512_{-0.0005}^{+0.0002}$&$3.6617_{-0.0437}^{+0.0293}$&2.078\\
52$^{a}$& $1.267_{-0.005}^{+0.011}$&$476.230_{-11.470}^{+10.870}$&$2.916_{-0.016}^{+0.014}$&$18.29_{-0.838}^{+0.832}$&$2.2675_{-0.0105}^{+0.0085}$&$0.1062_{-0.0004}^{+0.0006}$&$2.3737_{-0.0097}^{+0.0063}$&1.420\\
53$^{b}$& $1.454_{-0.015}^{+0.014}$&$354.260_{-20.625}^{+20.660}$&$2.867_{-0.014}^{+0.014}$&$80.55_{-3.452}^{+3.454}$&$6.6697_{-0.0287}^{+0.0223}$&$0.5388_{-0.0121}^{+0.0091}$&$7.1781_{-0.0241}^{+0.0199}$&0.949\\
54$^{a}$& $1.329_{-0.015}^{+0.021}$&$955.720_{-16.990}^{+17.230}$&$2.763_{-0.017}^{+0.018}$&$21.49_{-1.085}^{+1.102}$&$4.7125_{-0.0175}^{+0.0135}$&$0.2024_{-0.0006}^{+0.0003}$&$4.9149_{-0.0219}^{+0.0171}$&0.893\\
55$^{a}$& $1.338_{-0.014}^{+0.017}$&$1057.30_{-15.950}^{+15.430}$&$2.718_{-0.028}^{+0.026}$&$12.84_{-1.016}^{+0.983}$&$4.6489_{-0.0159}^{+0.0151}$&$0.1537_{-0.0007}^{+0.0003}$&$4.8025_{-0.0255}^{+0.0155}$&1.295\\
\hline
\end{tabular}
\noindent{
\leftline {   Fitted with $^{a}$ diskbb + power law, 
$^{b}$ diskbb + Gaussian + power law and $^{c}$ diskbb + compST + power law} 
\leftline {   $^{d}$ PCA 3-10 keV,  $^{e}$ PCA 10-25 keV \&   
$^{f}$ PCA 3-25 keV Model Fitted Photon Fluxes in number of photons$/cm^2/s$.}}
\end{table}

\begin{table}
\vskip -0.5cm
\scriptsize
\centering
\centerline {Table 3 (Cont'd)}
\centerline {PCA Spectral fitted parameters}
\vskip 0.2cm
\begin{tabular}{lccccccccc}
\hline
\hline
Obs.&diskbb&diskbb&Power-Law&Power-Law&Black Body&Power-Law&Total&$\chi^2_{red.}$\\
    &T$_{in}(keV)$&Norm.&Index($\Gamma$)&Norm.&Flux$^{d}$&Flux$^{e}$&Flux$^{f}$&\\
    (1)    &  (2)  & (3)  & (4)&   (5)   &  (6)    &    (7)      &  (8)     &  (9)   \\ 
\hline

101$^{a}$&$1.022_{-0.003}^{+0.009}$&$1953.60_{-31.170}^{+34.780}$&$1.396_{-0.016}^{+0.016}$&$0.016_{-0.007}^{+0.007}$&$1.7073_{-0.0063}^{+0.0067}$&$0.0091_{-0.0009}^{+0.0008}$&$1.7164_{-0.0084}^{+0.0046}$&1.349\\
102$^{a}$&$1.029_{-0.021}^{+0.020}$&$2034.60_{-28.110}^{+28.240}$&$1.646_{-0.024}^{+0.022}$&$0.035_{-0.008}^{+0.007}$&$1.8481_{-0.0071}^{+0.0079}$&$0.0099_{-0.0003}^{+0.0004}$&$1.8581_{-0.0101}^{+0.0059}$&1.910\\
103$^{a}$&$0.984_{-0.003}^{+0.004}$&$1228.10_{-33.400}^{+34.020}$&$1.759_{-0.150}^{+0.155}$&$0.026_{-0.008}^{+0.007}$&$0.8836_{-0.0076}^{+0.0057}$&$0.0042_{-0.0004}^{+0.0004}$&$0.8883_{-0.0076}^{+0.0067}$&1.557\\
104$^{a}$&$1.037_{-0.024}^{+0.027}$&$1948.50_{-29.920}^{+28.350}$&$1.045_{-0.020}^{+0.024}$&$0.025_{-0.004}^{+0.005}$&$1.8452_{-0.0092}^{+0.0078}$&$0.0092_{-0.0032}^{+0.0021}$&$1.8544_{-0.0114}^{+0.0066}$&1.173\\
105$^{a}$&$0.957_{-0.011}^{+0.018}$&$1357.70_{-105.50}^{+100.31}$&$2.168_{-0.022}^{+0.022}$&$0.033_{-0.004}^{+0.003}$&$0.8628_{-0.0028}^{+0.0046}$&$0.0042_{-0.0011}^{+0.0010}$&$0.8669_{-0.0066}^{+0.0059}$&1.690\\
\hline
\end{tabular}
\noindent{
\leftline {   Fitted with $^{a}$ diskbb + power law, 
$^{b}$ diskbb + Gaussian + power law and $^{c}$ diskbb + compST + power law} 
\leftline {   $^{d}$ PCA 3-10 keV,  $^{e}$ PCA 10-25 keV \&   
$^{f}$ PCA 3-25 keV Model Fitted Photon Fluxes in number of photons$/cm^2/s$.}}
\end{table}

\begin{table}
\scriptsize
\centering
\centerline {Table 4}
\centerline {F-test results for the 5 set of spectra of Fig.11}
\vskip 0.2cm
\begin{tabular}{lcccccc}
\hline
\hline
Obs. Id&Set 1$^*$& $\chi^2_{red.}$ \& d.o.f.&Set 2$^*$&$\chi^2_{red.}$ \& d.o.f.$^*$ &F-test Results$^\dagger$&Remarks\\
\hline
90704-04-01-00 &diskbb + po&4.009, 50&diskbb + compST + po&2.582, 47&8.690, 1.078e-04& \\
90704-04-01-00 &diskbb + po&4.009, 50&diskbb + ga + po&1.019, 48&70.438, 5.266e-15&Set 2 was used\\
\hline
91702-01-02-00G&diskbb + po&2.121, 50&diskbb + compST + po&1.085, 47&14.957, 5.723e-07& \\
91702-01-02-00G&diskbb + po&2.121, 50&diskbb + ga + po&0.875, 48&34.166, 5.928e-10&Set 2 was used\\
\hline
91702-01-08-00 &diskbb + po&3.621, 45&diskbb + compST + po&2.716, 33&0.916, 0.542& \\
91702-01-08-00 &diskbb + po&3.621, 45&diskbb + ga + po&2.210, 43&10.043, 2.636e-04&Set 2 was used\\
\hline
91702-01-57-00G&diskbb + po&2.778, 50&diskbb + compST + po&0.936, 47&30.811, 3.644e-11& \\
91702-01-57-00G&diskbb + po&2.778, 50&diskbb + ga + po&0.934, 48&49.804, 1.955e-12&Set 2 was used \\
\hline
91702-01-79-00 &diskbb + po&8.172, 50&diskbb + compST + po&1.179, 47&92.924, 8.981e-20&Set 2 was used\\
91702-01-79-00 &diskbb + po&8.172, 50&diskbb + ga + po&4.99, 49&51.358, 3.665e-09& \\
\hline
\end{tabular}
\noindent{
\leftline {$^*$diskbb: disk black body, po: power-law, ga: Gaussian,
compST: Sunyaev \& Titarchuk Comptonization model, d.o.f.: Degrees of Freedom.}
\leftline { $^\dagger$F-test was done between Set 1 and Set 2 model fitted results and
in the column F statistic value \& probability values were given.}}
\end{table}

\begin{table}
\scriptsize
\centering
\centerline {Table 5}
\centerline {Fitted data with two component flow model}
\vskip 0.2cm
\begin{tabular}{lcccc}
\hline
\hline
Date& MJD & Halo rate & Disk rate \\
\hline
Mar. 3, 2005 & 53432.7930& 0.015& 0.0100\\
Mar. 9, 2005 & 53438.0539& 0.004& 0.0400\\
Mar. 10, 2005& 53439.7400& 0.009& 0.0650\\
Mar. 11, 2005& 53440.6780& 0.0043& 0.5100\\
Mar. 12, 2005& 53441.5109& 0.0100& 0.7800\\
May 16, 2005& 53506.9531& 0.0550& 1.5640\\
May 20, 2005& 53510.0994& 0.0560& 1.7400\\
Sep. 16, 2005& 53629.3761& 0.0038& 0.0700\\
Sep. 19, 2005& 53632.4557& 0.0029& 0.0300\\
Sep. 22, 2005& 53635.4696& 0.0065& 0.0140\\
Sep. 29, 2005& 53641.0383& 0.009&0.0100\\
\hline
\end{tabular}
\end{table}
                                                                                                               
\noindent {\bf Acknowledgments} 
DD and SM acknowledge fellowships from CSIR (NET) and DST (FTYS) respectively which supported them to carry out
this work. The RXTE data used in this paper are from public archive of NASA/GSFC.  
Part of this work was carried out while DD and SKC were visiting Abdus Salam ICTP, Trieste.\\


{}

\end{document}